\documentclass[11pt,onecolumn,a4paper] {article}

\usepackage[cmex10]{amsmath}
\usepackage{amssymb, bm, algorithm}
\usepackage[table,xcdraw]{xcolor}
\usepackage{cite}
\usepackage{algorithm,algorithmicx, algpseudocode}
\usepackage{comment, color, balance}
\usepackage{tikz}
\usetikzlibrary{shapes,arrows}
\usepackage[tmargin = 0.8in, lmargin = 0.8in, rmargin = 0.8in, bmargin = 0.8in]{geometry}

\def \0{\bm{0}}
\def \1{\bm{1}}

\def \c{\bm{c}}

\def \q{\bm{q}}

\def \s{\bm{s}}
\def \u{\bm{u}}
\def \v{\bm{v}}

\def \x{\bm{x}}

\def \C{\mathbf{C}}

\def \P{\mathbf{P}}
\def \Q{\mathbf{Q}}
\def \R{\mathbf{R}}
\def \T{\mathsf{T}}
\def \Her{\mathsf{H}}

\def \W{\mathbf{W}}

\newcommand*{\Real}[1]{\mathsf{Re} \left\{ #1 \right\} }
\newcommand*{\Imag}[1]{\mathsf{Im} \left\{ #1 \right\} }

\def \LambdaB{\mathbf{\Lambda}}

\def \ph{{\textstyle \frac{2\pi}{3}}}

\def \eqdef{\stackrel{\text{def}}{=}}

\def \domC{\mathbb{C}}
\def \domR{\mathbb{R}}

\usepackage{tikz}
\usetikzlibrary{shapes,arrows}
\usepackage{amsthm}
\newtheorem{remark}{Remark}

\usepackage{graphicx}
\usepackage{epstopdf}

\graphicspath{{clarke_pics/}}

\newcommand*{\minus}{%
  \leavevmode
  \hphantom{0}%
  \llap{%
    \settowidth{\dimen0 }{$0$}%
    \resizebox{1.1\dimen0 }{\height}{$-$}%
  }%
}

\begin{document}
\title{\textbf{A Data Analytics Perspective of the Clarke and Related Transforms in Power Grid Analysis}\\Teaching Old Power Systems New Tricks }
\date{}
\author{Danilo P.~Mandic, Sithan~Kanna, Yili~Xia, Ahmad~Moniri, and Anthony~G.~Constantinides}

%
             

%

\maketitle

%


\section*{}\vspace{-12mm}
Affordable and reliable electric power is fundamental to modern society and economy, with the Smart Grid becoming an increasingly important factor in power generation and distribution. In order to fully exploit it advantages, the analysis of modern Smart Grid requires close collaboration and convergence between power engineers and signal processing and machine learning experts.  Current  analysis techniques are typically derived from a Circuit Theory perspective; such an approach is adequate for only fully balanced systems operating at nominal conditions and non--obvious for data scientists -- this is prohibitive for the analysis of dynamically unbalanced smart grids, where Data Analytics is not only well suited but also necessary. A common language that bridges the gap between Circuit Theory and Data Analytics, and the respective community of experts, would be a natural step forward. To this end, we revisit the Clarke and related transforms from a subspace, latent component, and spatial frequency analysis frameworks, to establish fundamental relationships between the standard three--phase transforms and modern Data Analytics. We show that the Clarke transform admits a physical interpretation as a ``spatial dimensionality'' reduction technique which is equivalent to Principal Component Analysis (PCA) for balanced systems, but is sub--optimal for dynamically unbalanced systems, such as the Smart Grid, while the related Park transform performs further ``temporal'' dimensionality reduction. Such a perspective opens numerous new avenues for the use  Signal Processing and Machine Learning in power grid research, and paves the way for innovative  optimisation, transformation, and analysis techniques that are not accessible to arrive at from the standard Circuit Theory principles, as demonstrated in this work through the possibility of  simultaneous frequency estimation and fault detection via adaptive Clarke and Park transforms. In addition, the introduced seamless transition between the Circuit Theory concepts and Data Analytics ideas  promises to provide a straightforward and unifying platform for further the understanding of sources of imbalance in modern power grids, together with an avenue for learning strategies, optimal parameter selection, and enhanced interpretation of Smart Grid problems and  new avenues for the mitigation of these issues.
  In addition, the material may be useful in lecture courses in multidisciplinary research from Smart Grid to Big Data, or indeed, as interesting reading for the intellectually curious and generally knowledgeable reader.\vspace{2mm}\\
\begin{tabular}{|p{0.95\columnwidth}|}
\rowcolor[HTML]{BBDAFF} \hline
\vspace{0.05cm} \bf Tribute to Edith Clarke, a pioneer of power grid analysis \vspace{0.1cm}                \\
 \vspace{0.5mm}
Edith Clarke (1883-1959) is a true pioneer in the application of circuit theory and mathematical techniques to electrical power systems. She was the first woman to obtain an M.S. in Electrical Engineering from MIT, in 1919, and the first female professor of Electrical Engineering in the USA, having been appointed at the University of Texas at Austin, in 1947. Her pivotal contributions were concerned with the development of algorithms for the simplification of the laborious computations involved in the design and operation of electrical power systems \cite{Brittain_Clarke_85}. One of her early inventions was the Clarke calculator (1921), a graphical device that solved power system equations 10 times faster than a human computer {\cite{E_Clarke:Human_Computer}}. The Clarke transform, also known as the $\alpha \beta$ transform, was introduced by Edith Clarke in 1943, and has since been established as a  fundamental and indispensable tool for the analysis of three--phase power systems. With the advent of Smart Grid, the Clarke transform represents an underpinning technology  for signal processing, control and machine learning  applications related state estimation, frequency tracking, and fault detection  \cite{Cantelli_ClarkePID_06,Barbosa_FuzzyClarke_2011,Yili_GridSPM_2012},   the most important aspects in the development of the future Smart Grid.\\ \hline 
\end{tabular}

\section*{Challenges in Smart Grid: A Fertile Ground for Data Analytics}
There is substantial interest in transforming the way we both produce and use energy as
current ways are not sustainable. For the electrical power grid this involves fundamental
paradigm shifts as we build a smart grid, adopt more renewable energy sources, and
promote more energy efficient practices. A smart grid delivers electricity from suppliers
to users using digital technology and has a number of properties including incorporating
all forms of energy generation and storage, using sensor information, enabling active
participation by end users, being secure and reliable, and using optimization and control
to make decisions; see for example, the Energy Independence and Security Act 2007, Section 1304 Smart Grid RD\&D
Highlights. This will require fundamental shifts in the way we analyse and design power systems and prominent involvement of modern Data Analytics disciplines which are currently outside the standard Power Systems,  such as those enabled by 
Signal Processing and Machine Learning. Although we have just begun to investigate a whole host of e.g. Signal Processing issues for the smart grid strategy, these new technologies will undoubtedly
be critical to the efficient use of limited and intermittent power resources in the future. The first and fundamental step in this direction is to bridge the gap between the Power Systems community and the Data Analytics communities by establishing  a common language for  the understanding and interpretation of system behaviour, the aim of this Perspective.\vspace{2mm}\\
{\bf Economic Value of the Smarter Grid.}
To depict the sheer scale of the required changes to current practice, we summarise the recent findings from {\cite{Arnold_Proc_IEEE_2011}}:\vspace{-2mm}
\begin{itemize}
\item The Brattle Group has estimated that the investment needed in replace old generation of power plants with the new ones would be about USD 560 billion by 2030;\vspace{-2mm}
\item  In the  USA, about 40\% of human-caused emissions of $CO_2$ are due to generation of electricity;\vspace{-2mm}
\item  If the power grids were just 5\% more efficient, the resultant energy and emission reductions would be equivalent to permanently eliminating 53 million cars;\vspace{-2mm}
\item Capacity to meet demand during the top 100 peak hours in the year accounts for 10-20\% of total electricity costs;\vspace{-2mm}
\item The cost of outages to the USA economy is about USD 80 billion annually.\vspace{-2mm}
\end{itemize}
Yet, current centralised power plants are at best 35\% efficient  while the renewable sources affect the stability and inertia in current power systems and are therefore not used to their full capacity.\vspace{2mm}\\
{\bf Power Quality Issues.} System frequency is the most important power quality parameter;  its rise indicates more generation than consumption while a decrease indicates less generation than consumption.  The IEEE 1547 Standard specifies  that  a distributed generation source must disconnect from a locally islanded system within 2 seconds; it also requires disconnecting for sagging voltage under high demand (voltage sags are described by the IEEE Standard 1159--1995). However, disconnecting a large number of local generators (e.g. solar) can cause the low-voltage condition to accelerate {\cite{Arnold_Proc_IEEE_2011}}, and can also affect the current way of estimating power quality parameters (frequency, voltage phasors). This all calls for accurate frequency estimator which are robust under unbalanced system conditions.

Current estimation in three--phase systems is routinely performed through the Clarke and related transforms, which are designed for stable grids operating in nominal conditions. However, smart grids introduce dynamically unbalanced conditions which yield  incorrect frequency estimates due to {\cite{Bollen_2001,Dora_Chat_2011,Bollen_DSP_Power_IEEE_SPM_2009}}:\vspace{-2mm}
\begin{itemize}
\item Inadequacy of the Clarke and related transforms for unbalanced system conditions;\vspace{-2mm}
\item False frequency estimates when the system is experiencing voltage sags, that is, off--nominal amplitudes and/or phases of the three phase voltages, even if the systems frequency remains at a nominal $\omega_{\circ} \in \{ 50 Hz, 60 Hz \}$;\vspace{-2mm}
\item Some loads (furnaces, cyclo-converters) introduce inter--harmonics that are not integer multiplies of the fundamental frequency. These cannot be estimated using spectral techniques and tend to drift over time thus affecting systems  prone to resonance (low damping or a high Q factor).	
\end{itemize}
{\bf Opportunities for Data Analytics Research.} Three-phase systems can be inherently difficult to analyse as the electrical quantities involved are coupled by design while also exhibiting redundancies.  During her early career as a human ``computer" with the General Electric company, Edith Clarke routinely faced problems related to the  simplification of the analyses of three-phase circuits. Fast forward a century, and three-phase systems pose another class of practical problems, essentially of a  signal processing and machine learning nature which include:\vspace{-2mm}
\begin{itemize}
\item In smart grids, the effects arising from the on--off switching of various subgrids and the dual roles of generators/loads  will produce transients and spurious frequency/phasor estimates; the analysis thus requires modern Signal Processing and Machine Learning techniques;\vspace{-2mm}
\item Accurate  change of frequency trackers and rapid frequency estimators are a pre--requisite for the operation of smart grid, but their design is beyond the remit of Power Systems engineering;\vspace{-2mm}
\item Rapid frequency trackers are envisaged to be part of many appliances, as in Smart Grid  we not only must dynamically bring in new generators and interconnect the grid, but also  smart loads must be able to detect rapid frequency changes and take action;\vspace{-2mm}
\item Loss of mains detection from voltage dips and off--nominal frequencies is critical for system balance, these imbalances will be much more prominent in low--inertia grids of the future;\vspace{-2mm}
\item Current analyses compute features over a predefined time interval, such as 200ms, 3s, 1m, 10m, 2h. These are adequate for standard grids with power reserve and high interia (International Standard IEC 6100-4-30 ).    For example, the PQ variations are currently calculated over a 200ms window -- too coarse for rapid and real--time monitoring and analysis in smart grids where the required time scales are in the region of 2ms and below.\vspace{-2mm}
\end{itemize}
All in all,  it is critical that  frequency/phasor estimator remain accurate during the various interconnections, transients, faults, and voltage sags (IEEE Standard 1159--1995), while at the same time having intelligence to indicate whether the system experienced 1--, 2-- or 3--phase fault; this ``smart frequency'' area has been subject of some recent patents {\cite{DPM_et_al:Frequency_Estimation}} and ongoing research {\cite{Tirza_Imbalance_Symchorphasor_IEEE_TPS_2018,Routtenberg_KCL_2013,Pradhan_LMS_2005,Routray_EKFSLAR2_2002,Kanna_Stabilising_Clarke_El_Letters_2017, Huang_SE_SPM2012, Yili_GridSPM_2012}}.
\section*{Sources of Redundancy in Power System Analysis}
We shall start by investigating the redundancy of information--bearing signals in three-phase systems, in order to establish a link between the current Circuit Theory inspired dimensionality reduction techniques and a more general  Latent Component Analysis (LCA) view rooted in Data Analytics.

\subsection*{Exploiting Redundancy in Three-Phase Signal Representation}
\begin{figure}[htbp] \centering
\definecolor{myblue}{rgb}{0, 0.4, 0.8}
\definecolor{myred}{rgb}{0.75, 0, 0}
\definecolor{mygreen}{rgb}{0.2, 0.7, 0}
\definecolor{myorange}{rgb}{0.9,0.5,0.2}

\begin{tikzpicture}[>=latex, xscale=0.65,yscale=0.65]

\node[align=left] at (2,3) {Three-phase frame};
\node[align=left] at (10,3) {Clarke basis};
\node[align=left] at (18,3) {Park basis};

\draw[style=help lines] (0,0) (3,2);

\coordinate (vec1) at (120:1.95); 
\coordinate (vec2) at (240:1.95);
\coordinate (vec3) at (0:1.95);

\coordinate (vec4) at (90:1.95);
\coordinate (vec5) at (40:1.95);
\coordinate (vec6) at (130:1.95);

\draw[->,ultra thick,myblue] (1.5,0) -- ++(vec1) node[right] {$v_c$};
\draw[->,ultra thick,myblue] (1.5,0) -- ++(vec2) node[right] {$v_b$};
\draw[->,ultra thick,myblue] (1.5,0) -- ++(vec3) node [below] {$v_a$};

\draw[->,thick] (4,0) -- ++(4.45,0) node[] {};

\node[align=center, text width =3cm] at (6,-1.1) {Spatial \\ Dimensionality \\ Reduction
};

\draw[->,ultra thick,myred] (9,-1) -- ++(vec3) node[right] {$v_{\alpha}$};
\draw[->,ultra thick,myred] (9,-1) -- ++(vec4) node[above] {$v_{\beta}$};

\draw[->,thick] (12,0) -- ++(4.45,0) node[] {};

\node[align=center, text width =3cm] at (14.2,-1.1) {Temporal \\ Dimensionality \\ Reduction
};

\draw[->,ultra thick,mygreen] (18,-1) -- ++(vec5) node[right] {$v_{d}$};
\draw[->,ultra thick,mygreen] (18,-1) -- ++(vec6) node[above] {$v_{q}$};
\draw[ultra thick,black, dashed] (18,-1) -- ++(vec3) node[] {};

\node[align=left] at (19.7,-0.6) {$\theta_k$};

\draw [thick] (19,-1) arc [start angle=0, end angle=40, radius=1]
    node [below] {};
\end{tikzpicture}

\vspace{7mm}
\begin{tikzpicture}[xscale=0.6,yscale=0.6]
  \draw[->] (0,-1.3) -- (0,1.3);
  \draw[->] (0,0)--(5.8,0);
  \draw[domain=0:5.12,samples=100,myblue, ultra thick] plot(\x,{cos(\x r)});
  \draw[domain=0:5.12,samples=100,myblue, ultra thick, dashed] plot(\x,{cos((\x-2.09) r)});
  \draw[domain=0:5.12,samples=100,myblue, ultra thick, densely dotted] plot(\x,{cos((\x +2.09) r)});
  
  \node[align=left, myblue] at (-0.5,1) {$v_{a}$};
    \node[align=left, myblue] at (-0.5, -0.3) {$v_{b}$};

  \node[align=left, myblue] at (-0.5, -1) {$v_{c}$};

    \draw[->] (7.8,-1.3) -- (7.8,1.3);
  \draw[->] (7.8,0)--(14.6,0);
  \draw[domain=7.8:14.92-1,samples=100,myred, ultra thick] plot(\x,{cos((\x-8.8) r)});
  \draw[domain=7.8:14.92-1,samples=100,myred, ultra thick,densely  dashed] plot(\x,{sin((\x-8.8) r)});

  \node[align=left, myred] at (7.2,0.7) {$v_{\alpha}$};
	\node[align=left, myred] at (7.2, -1) {$v_{\beta}$};

      \draw[->] (16.6,-1.3) -- (16.6,1.3);
  \draw[->] (16.6,0)--(22.4,0);
  \draw[mygreen, ultra thick] (16.6,1)--(21.72,1);
    \draw[mygreen, ultra thick, densely dashed] (16.6,0.2)--(21.72,0.2);
  
  \node[align=left, mygreen] at (16,1) {$v_q$};
\node[align=left, mygreen] at (16, 0) {$v_d$};

\end{tikzpicture}
\caption{Geometric interpretation of the Clarke and Park Transforms through the corresponding ``spatial'' and ``temporal'' dimensionality reductions.}
\label{fig:frames}

\end{figure}
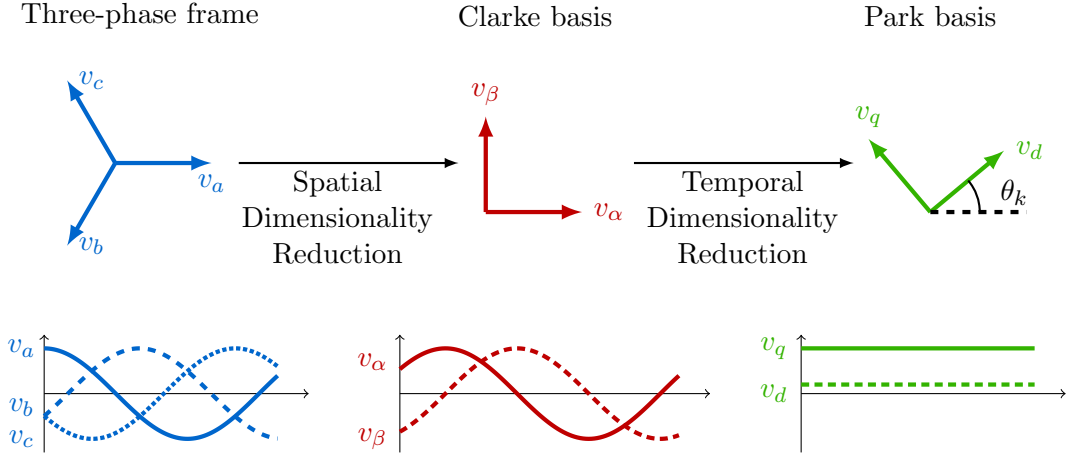

Consider a sampled three-phase voltage measurement vector, $\s_k$, which at a discrete time instant $k$, is given by 
\begin{align} \label{eq:3_phase}
\s_k = \begin{bmatrix}
 v_{a,k}   \\
  v_{b,k}  \\
   v_{c,k} 
\end{bmatrix} =  \begin{bmatrix}
 V_{a}\cos(\omega k + \phi_{a})   \\
 V_{b}\cos(\omega k + \phi_{b} - \ph)  \\
 V_{c}\cos(\omega k + \phi_{c} +\ph) 
\end{bmatrix} , 
\end{align}
where $V_{a}, V_{b}, V_c$ are the amplitudes  of the phase voltages $v_{a,k}$,  $v_{b,k}$,  $v_{c,k}$,  while $\omega = 2\pi f T$ is the fundamental angular frequency,  with $f$ the fundamental power system frequency and $T$  the sampling interval. The phase values for phase voltages are denoted by $\phi_{a}, \phi_{b}$, and $\phi_c$.  
\begin{remark}\label{Rem:Balanced_System}
 The three-phase power system is considered to be in a balanced condition if
 \begin{enumerate}
 \item The magnitudes of the phase voltages in \eqref{eq:3_phase} are equal, that is, $V_a = V_b = V_c$, 
 \item The phase angle separation between the phase voltages is uniform and equal to $\ph$, that is, $\phi_a= \phi_b=\phi_c$, across the phase voltages.
\end{enumerate}
\end{remark}
Early power engineers were able to effectively reduce the dimensionality in representing the three-phase signal in \eqref{eq:3_phase} {\bf by changing the reference frame (or basis) of the three--phase power voltage signal}, the so--called voltage transformations \cite{Microsemi}.   Figure \ref{fig:frames} illustrates effects of the three-phase transformations considered in this paper -- the Clarke Transform and the closely related Park Transform.  
\begin{remark}\label{Rem:Clarke_Park_Meaning} 
 Figure \ref{fig:frames} allows us to provide a modern interpretation of the operation of the Clarke and Park transforms, whereby the Clarke Transform reduces the three--dimensional ``spatial information space'' in three-phase power signals to the two--dimensional $\alpha \beta$ space, while the Park transform applies a two--dimensional time-varying basis to the Clarke transform, in the form of a rotation matrix whose bases rotate at the fundamental power system frequency of $50$ Hz, to further reduces the ``temporal information space'' to  only two constants, $v_{d}$ and $v_{q}$.   
\end{remark}

\subsection*{Signal Processing View of Spatial Redundancy in Three--Phase Power Systems}
We now show that the three-phase power signal in \eqref{eq:3_phase} is essentially over-parametrised, thus paving the way for a Data Analytics perspective of the Clarke Transform. To this end, consider the empirical covariance matrix of the three-phase voltage signal, $\s_k $ in \eqref{eq:3_phase}, defined as $\mathsf{cov}(\s_k) \eqdef \R_{\s}$, which can be computed from $N$ consecutive samples of  $\s_k $  as
\begin{align} 
\label{eq:R_def}
\R_{\s} =  {} &  \lim_{N \rightarrow \infty} \frac{1}{N}\sum_{k = 0}^{N-1}  \s_k\s_k^\Her, 
\end{align}
where the symbol $(\cdot)^\Her$ denotes the Hermitian transpose operator. 

\noindent {\bf Phasor representation.} From  the three-phase voltage, $\s_k$ in \eqref{eq:3_phase}, upon employing the identity $\cos(x) = (e^{jx} + e^{-jx})/{2}$, we arrive at its phasor representation in the form 
%
%
%
%
%
\begin{align} \label{eq:3_phase_complex}
\s_k = \frac{1}{2}\begin{bmatrix}
\bar{V}_a      \\
 \bar{V}_b      \\
\bar{V}_c 
\end{bmatrix} 
e^{j\omega k }   +  \frac{1}{2}  \begin{bmatrix}
 \bar{V}^*_a      \\
 \bar{V}^*_b      \\
 \bar{V}^*_c
\end{bmatrix}  e^{-j\omega k } 
\end{align}
where, for compactness,  the time-independent phasors, 
$\bar{V}_a = \frac{V_a}{\sqrt{2}} e^{j\phi_a}$, $\bar{V}_b = \frac{V_b}{\sqrt{2}} e^{j\left(\phi_b -  \ph\right)}$ 
and $\bar{V}_c = \frac{V_c}{\sqrt{2}} e^{j\left(\phi_c +  \ph\right)}$, can be comprised into the phasor vector 
\begin{align}\label{eq:phasor_def_ori}
\v \eqdef{} 
\begin{bmatrix}
\bar{V}_a, &\bar{V}_b, &  \bar{V}_c
\end{bmatrix}^\T.
\end{align}
Without loss of generality, we shall consider normalised versions of the phasors  (relative to $\bar{V}_a$), and define $\delta_{i} \eqdef \bar{V}_{i}/\bar{V}_a $, $i \in \{a, b, c \}$,  with $\delta_a = 1$, to give  
%
%
%
%
%
\begin{align} \label{eq:s_k}
\s_k = \frac{1}{2} \left(\v e^{j\omega k} + \v^* e^{-j\omega k}\right)
\end{align}
%
%
so that the normalised version of the phasor vector,  $\v$ in \eqref{eq:phasor_def_ori}, now becomes 
\begin{align} \label{eq:phasor_vec}
\v = \begin{bmatrix}
1,   & \delta_b,  & \delta_c 
\end{bmatrix}^\T. 
\end{align}
In order to arrive at the final expression for the empirical covariance matrix, $\R_{\s}$ in \eqref{eq:R_def},  observe from  \eqref{eq:s_k} that the individual outer products, $\s_k\s_k^\Her$ in  \eqref{eq:R_def}, represent an average of four outer products, that is
\begin{align} \label{eq:sk_outer}
\s_k\s_k^\Her = \frac{1}{4}\left( \v\v^{\Her} + \v^*\v^\T + \v \v^\T e^{2j\omega k} + \v^* \v^{\Her} e^{-2j\omega k} \right). 
\end{align}
For $\omega \ne 0$ or $\omega \ne \pi$,  and for a large enough $N$, the following holds \cite[p. 56]{Kay93}
\begin{align} 
\lim_{N \rightarrow \infty} \frac{1}{N}\sum_{k = 0}^{N} e^{\pm 2j\omega k} = 0,
\end{align}
so that the last two outer products in \eqref{eq:sk_outer} vanish and the individual outer product within the covariance matrix for a  general 3-phase power voltage measurement become
\begin{align} \label{eq:R_3}
\s_k\s_k^\Her  =  &  \frac{1}{4}\left( \v\v^{\Her} + \v^*\v^\T  \right) = \frac{1}{2}\Real{\v\v^\Her} = \frac{1}{2}\left( \v_r \v_r^{\T} + \v_i \v_i^\T  \right), 
\end{align}
where $\v_r = \Real{\v}$ and $\v_i = \Imag{\v}$ denote the real and imaginary part of the phasor vector $\v$ defined in \eqref{eq:phasor_vec}.
\begin{remark}\label{rem:rank}
Observe from \eqref{eq:R_3}  that the $3 \times 3$ covariance matrix, $\R_{\s}$ in \eqref{eq:R_def}, of the trivariate  three--phase voltage signal, $\s_k$, is rank-deficient (Rank--2) as it represents a sum of two Rank-1 outer products, $\v_r\v_r^\T$ and $\v_i\v_i^\T$.  In other words, without loss in information the three-phase signal in \eqref{eq:3_phase_complex}  can be projected onto a two--dimensional subspace spanned by $\v_r\v_r^\T$ and $\v_i\v_i^\T$. This implies that the use of  all three data channels (system phases) is  redundant in the analysis, and  offers a Data Analytics justification for the  Clarke Transform. 
\end{remark}
We next proceed with the formal definition of the Clarke Transform, and show that its dimensionality reduction principle admits a Principal Component Analysis (PCA) interpretation.   
\section*{Clarke Transform -- A Fundamental Tool in Power System Analysis}
%
%
%
The Clarke transform, also known as the $\alpha\beta$ transform, was introduced from a Circuit Theory viewpoint and aims to  change the basis of the original vector space where the three-phase signal $\s_k$ in \eqref{eq:3_phase}  resides, to a basis defined  by the columns of  the so-called Clarke matrix, to yield the Clarke--transformed $ v_{0, k}, v_{\alpha, k}, v_{\beta, k}$  voltages in the form 
\begin{align}	 \label{eq:clarke_1}			
	\begin{bmatrix} v_{0,k} \\ v_{\alpha,k} \\  v_{\beta,k}\end{bmatrix} = {} &		\underbrace{\sqrt{\frac{2}{3}} \begin{bmatrix} \frac{\sqrt{2}}{2} &  \frac{\sqrt{2}}{2} & \frac{\sqrt{2}}{2} \\ 1 & -\frac{1}{2} & -\frac{1}{2} \\ 0 & \frac{\sqrt{3}}{2} & -\frac{\sqrt{3}}{2} \end{bmatrix}}_{\text{Clarke matrix}}
 \underbrace{\begin{bmatrix} v_{a,k} \\ v_{b,k} \\ v_{c,k}\end{bmatrix}}_{\displaystyle \s_k}, 
\end{align}
The quantities $v_{\alpha,k}$ and   $v_{\beta,k}$ are referred to as the $\alpha$ and $\beta$ sequences, while the term   $v_{0,k}$ is called the zero-sequence, as it is null when the three-phase signal $\s_k$ is balanced (see Remark \ref{Rem:Balanced_System}).

\begin{figure*}[htbp]
\centering
    {\includegraphics[scale=0.36]{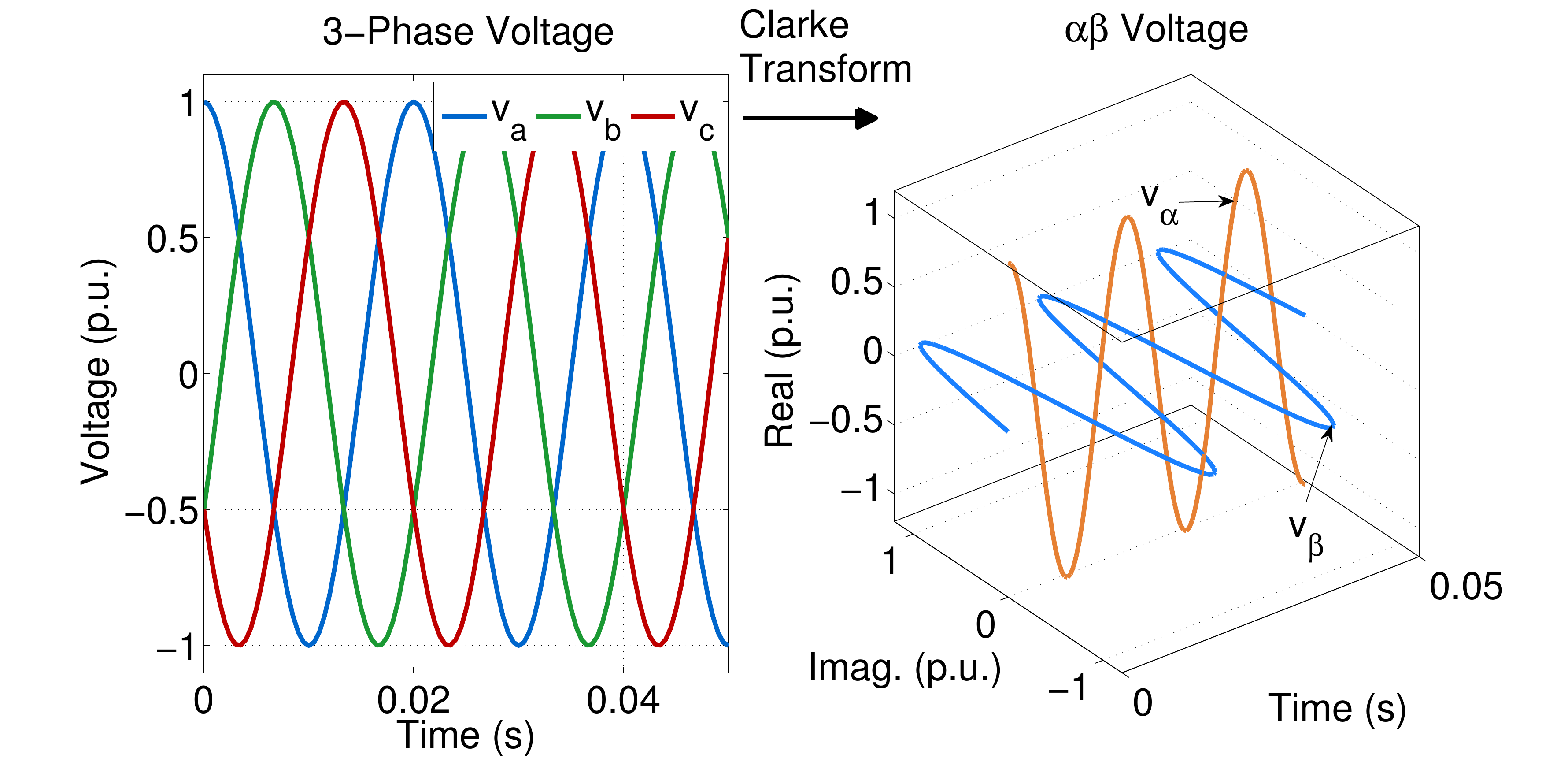}} 
  \caption{Waveforms of Clarke--transformed three--phase voltages. The Clarke voltages $v_{\alpha}$ and $v_{\beta}$ are orthogonal and admit a convenient complex valued representation in the form $s_{i,k} = v_{\alpha,i, k} + jv_{\beta,i, k}$.}
  \label{fig:Clarke_pic}
\end{figure*}

\begin{remark}\label{Rem:Standard_Clarke} 
The traditional power grid is typically in a balanced condition due to its huge inertia, and therefore,  only $v_{\alpha, k}$ and $v_{\beta, k}$ are used in its analysis since balanced phase voltages yield $ v_{0,k}=0 $. The ``standard'' version of the Clarke transform thus employs only the last two rows of the Clarke matrix in \eqref{eq:clarke_1}, to project the three-phase voltage in \eqref{eq:3_phase} onto a 2D subspace spanned by these columns, that is
\begin{align}	 \label{eq:clarke_red}			
	\begin{bmatrix}  v_{\alpha,k} \\  v_{\beta,k}\end{bmatrix} = {} &		\underbrace{\sqrt{\frac{2}{3}} \begin{bmatrix} 1 & -\frac{1}{2} & -\frac{1}{2} \\ 0 & \frac{\sqrt{3}}{2} & -\frac{\sqrt{3}}{2} \end{bmatrix}}_{\text{Reduced Clarke matrix:}\: \displaystyle \C}
 \begin{bmatrix} v_{a,k} \\ v_{b,k} \\ v_{c,k}\end{bmatrix}.
\end{align}
\end{remark}
This is further visualised in Figure \ref{fig:Clarke_pic} which provides a geometric interpretation of the Clarke transform for  balanced power systems.  Observe the mutually orthogonal nature of the $v_{\alpha, k}$ and $v_{\beta, k}$ components, which allows for their convenient combination into a complex-valued voltage, $s_{k} = v_{\alpha, k} + jv_{\beta, k}$. 

\subsection*{Park Transform}
The Park transform (also known as the $dq$ transform) is closely related to Clarke Transform and projects the three-phase signal $\s_k$ onto an orthogonal, time-varying frame which,  by virtue of rotating at the fundamental power system  frequency $\omega_{\circ}$ (50 Hz or 60 Hz), yields stationary constant outputs, $v_{d,k}, v_{q,k}$. In other words, the Park voltages $v_{d,k}, v_{q,k}$ are obtained from the Clarke's $\alpha\beta$ voltages in \eqref{eq:clarke_red} using a time-varying transformation given by \cite{Park_1929}
\begin{align}	 \label{eq:Park} 
	\begin{bmatrix} v_{d,k} \\  v_{q,k}\end{bmatrix} = {} &		\underbrace{ \begin{bmatrix}  \cos(\theta_k) & \sin(\theta_k)  \\    \minus \sin(\theta_k)   & \cos(\theta_k)  \end{bmatrix}}_{\text{Park Matrix:}\: \displaystyle \P_{\theta}} \begin{bmatrix}  v_{\alpha,k} \\ v_{\beta,k}\end{bmatrix}	.	
\end{align}
where $\theta_k = \omega_{\circ} k$, while the orthogonal direct and quadrature components, $v_{d,k}$ and $v_{q, k}$, can be combined  into a complex variable $v_k = v_{d, k} + jv_{q, k}$.  
\begin{remark}\label{Rem:Park_Transfork}
From the modern Data Analytics perspective, the Park matrix, ${\bf P}_{\theta}$, is a full-rank and time-varying clock--wise rotation matrix, with the determinant $\det({\bf P}_{\theta})=1$ and the unit--norm eigenvalues $|\lambda_{1,2}|=1$. It therefore does not amplify the original Clarke vector $[v_{\alpha,k}, v_{\beta,k}]^{T}$ but  only rotates it,  with the speed of rotation equal to the fundamental frequency of the power system, $\omega_{\circ}$.  
\end{remark}

\begin{figure}[htbp]
\centering
{\includegraphics[scale=0.42, clip=true, trim = 5mm 100mm 10mm 90mm ]{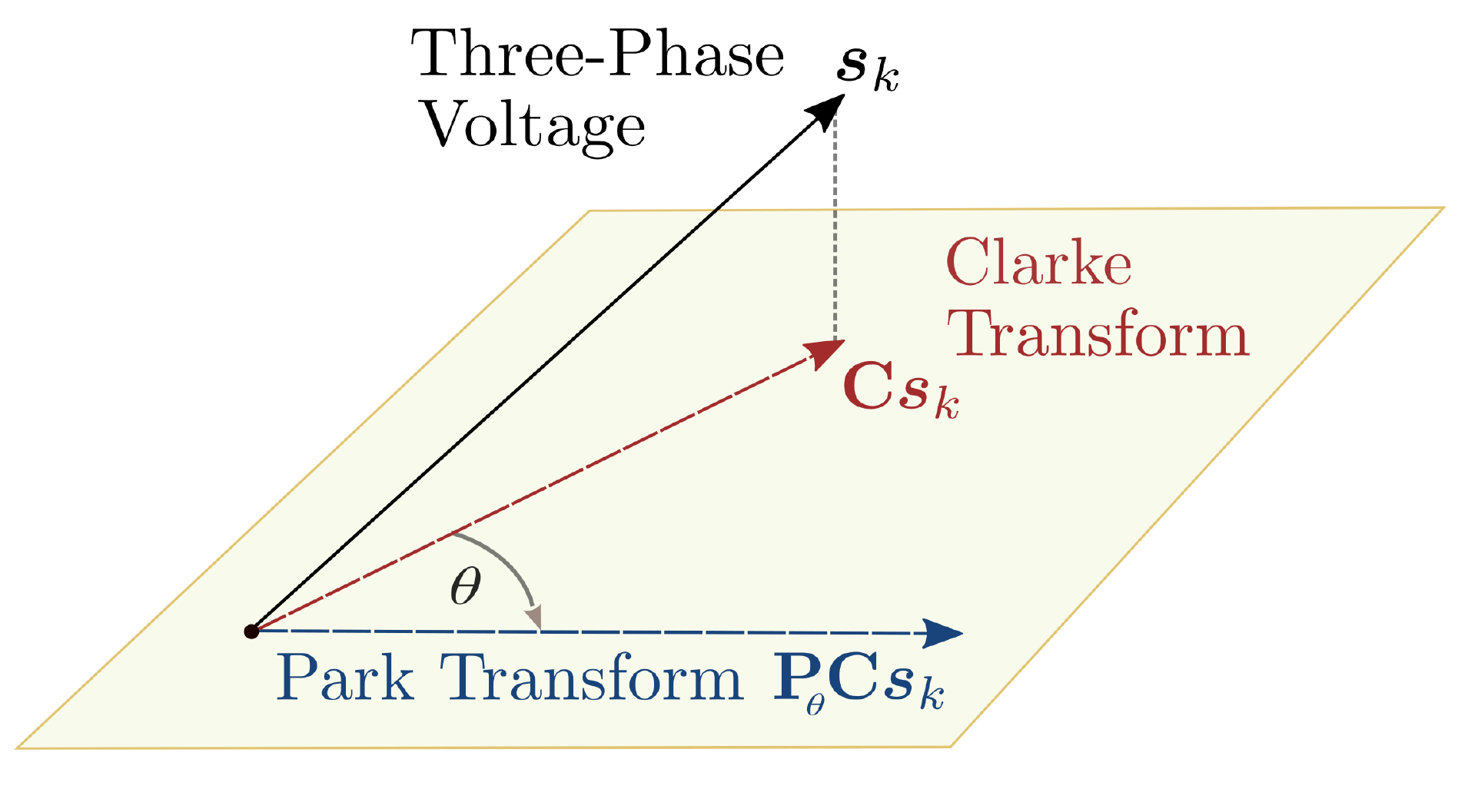}}
\caption{Geometric representation of the Clarke and Park three-phase transforms, applied to a three-phase voltage vector, $\s_k$.}
\label{fig:MatToVec}
\end{figure}
Figure \ref{fig:MatToVec} offers a geometric interpretation of the Clarke and Park transform of the three-phase voltage vector $\s_k$. Observe that,  while the Clarke transform matrix, $\C$ in \eqref{eq:clarke_red}  projects a 3D vector,  ${\bf s}_{k}$, onto a two-dimensional space spanned by its columns, the Park transform matrix, $\P_{\theta}$ in \eqref{eq:Park}, is a time-varying two--dimensionl rotation matrix  which changes only the direction of the 2-dimensional Clarke vector $ [v_{\alpha,k},v_{\beta,k}]^{T}=\C\s_k$.
\subsection*{Principal Component Analysis (PCA)}
Modern Data Analytics often employs Principal Component Analysis, in order to either separate meaningful data from noise, or to reduce the dimensionality of the original signal space while maintaining the most important information--bearing latent components in data. Consider a general data vector, $\x_k \in \domR^{M \times 1}$, for which the covariance matrix is defined as  
\begin{align}\label{eq:covX}
\text{cov}(\x_k) \eqdef \R_{\x} = \lim_{N \rightarrow \infty}\frac{1}{N} \sum_{k = 0}^{N-1} \x_k \x^\T_k. 
\end{align}
Then, this symmetric  covariance matrix $\R_{\x}$ admits the following eigenvalue decomposition 
\begin{align} \label{eq:eigenR}
\Q^\T\R_{\x}\Q =  \LambdaB 
\end{align}
where the diagonal eigenvalue matrix,  $\LambdaB  = \mathsf{diag}\{\lambda_1,\lambda_2, \ldots, \lambda_M   \} $, indicates the power of each component within $\x_k$, while the matrix of eigenvectors, $\Q_{r} = [\q_1, \q_2, \ldots, \q_M]$,   designates the principal  directions in the data.  

Suppose the signal $\x_k$ is to be transformed into a vector, $\u_k \in  \domR^{M \times 1}$, with  the same dimensionality as the original signal $\x_k$,  using a linear transformation matrix $\W$, to give 
\begin{align}\label{eq:KLT_cond}
\u_k = \W\x_k, \quad \text{where} \quad \mathsf{cov}(\u_k) = \LambdaB. 
\end{align}
The principal component analysis (PCA), also known as the Karhunen-Loeve transform, states that  the above  transformation matrix, $\W$, can be obtained from the eigenvector and eigenvalue matrices in  \eqref{eq:eigenR} as    $\W = \Q^{\T}$ \cite{Shlens_PCA2014}. In other words,
\begin{align}\nonumber 
 \mathsf{cov}(\u_k) = {} &   \lim_{N \rightarrow \infty}\frac{1}{N} \sum_{k = 0}^{N-1} \u_k \u^\T_k  \\ \nonumber
 = {} &  \W\left(\lim_{N \rightarrow \infty}\frac{1}{N} \sum_{k = 0}^{N-1} \x_k \x^\T_k \right) \W^\T \\ \label{eq:covs}
 = {} & \Q^{\T}\R_{\x} \Q = \LambdaB 
\end{align}
This formulation admits a convenient dimensionality reduction by retaining only $r < M$ largest  eigenvalues and the corresponding eigenvectors of $\R_{\x}$. The so obtained transformed data vector, $\u_{r,k} \in \domR^{r \times 1}$, is of dimension $r < M$  and is given by  
\begin{align} \label{eq:PCA_red}
\u_{r, k} = \Q^\T_{1:r} \x_k
\end{align}
where $\Q_{1:r} = [\q_1, \q_2, \ldots, \q_r]$, while $r$ stands for  the $r$ largest eigenvalues in $\LambdaB$. In other words, the PCA-based dimensionality reduction scheme in \eqref{eq:PCA_red} selects the   directions in which the data expresses  maximum variance, designated by the directions of the principal eigenvectors of the data covariance matrix, $\R_{\x}$. 
\section*{Clarke Transform as a Principal Component Analyser}
We have seen that for a balanced power system, the phasor vector, $\v$ in \eqref{eq:phasor_vec}, takes the form   
\begin{align}\label{eq:v_bal}
	\v = \begin{bmatrix} 1,  & e^{-j\ph}, & e^{j\ph} \end{bmatrix}^\T 
\end{align}
so that the covariance matrix of the  normalised three--phase power signal, $\s_k$, now becomes
\begin{align} 
\R_{\s} = {} & \frac{1}{2}\Real{\v\v^\Her} = 
			\frac{1}{4} \begin{bmatrix}
		2 & \minus 1 & \minus 1 \\ 
		\minus 1  & 2 & \minus 1  \\
		\minus 1  & \minus 1  &2  		
		 \end{bmatrix}.			 \label{eq:R3}				
\end{align} 
and thus admits the  eigen-decomposition in \eqref{eq:eigenR}, to yield
\begin{align}\label{eq:R_s}
\R_{\s}= {} & \Q  \mathbf{\Lambda} \Q^\T. 
\end{align}
By inspection of  $\R_{\s}$ in \eqref{eq:R3}, from the first eigenvector-eigenvalue pair, $(\q_1, \lambda_1)$, we have  
\begin{align}
\R_{\s}\q_1 = \mathbf{0} \implies \q_1 = \frac{1}{\sqrt{3}}\1,\: \lambda_1 = 0.
\end{align}
To find the remaining eigenvector-eigenvalue pairs, consider again the outer products within the covariance matrix, given in \eqref{eq:R_3}, and the normalised phasor vector, $\v$ in  \eqref{eq:v_bal}.  Notice that its real part, $\v_r = \Real{\v} = [1,   -\frac{1}{2},    -\frac{1}{2}]^\T$, and its imaginary part, $\v_i = \Imag{\v} = [0,   -\frac{\sqrt{3}}{2},    \frac{\sqrt{3}}{2}]^\T$,  are  orthogonal,  that is,  $\v_r^\T\v_i = 0$. 

\begin{figure}[htbp]  \centering 
     {\includegraphics[clip = true, trim =0mm 0mm 0mm 10mm, scale=0.47]{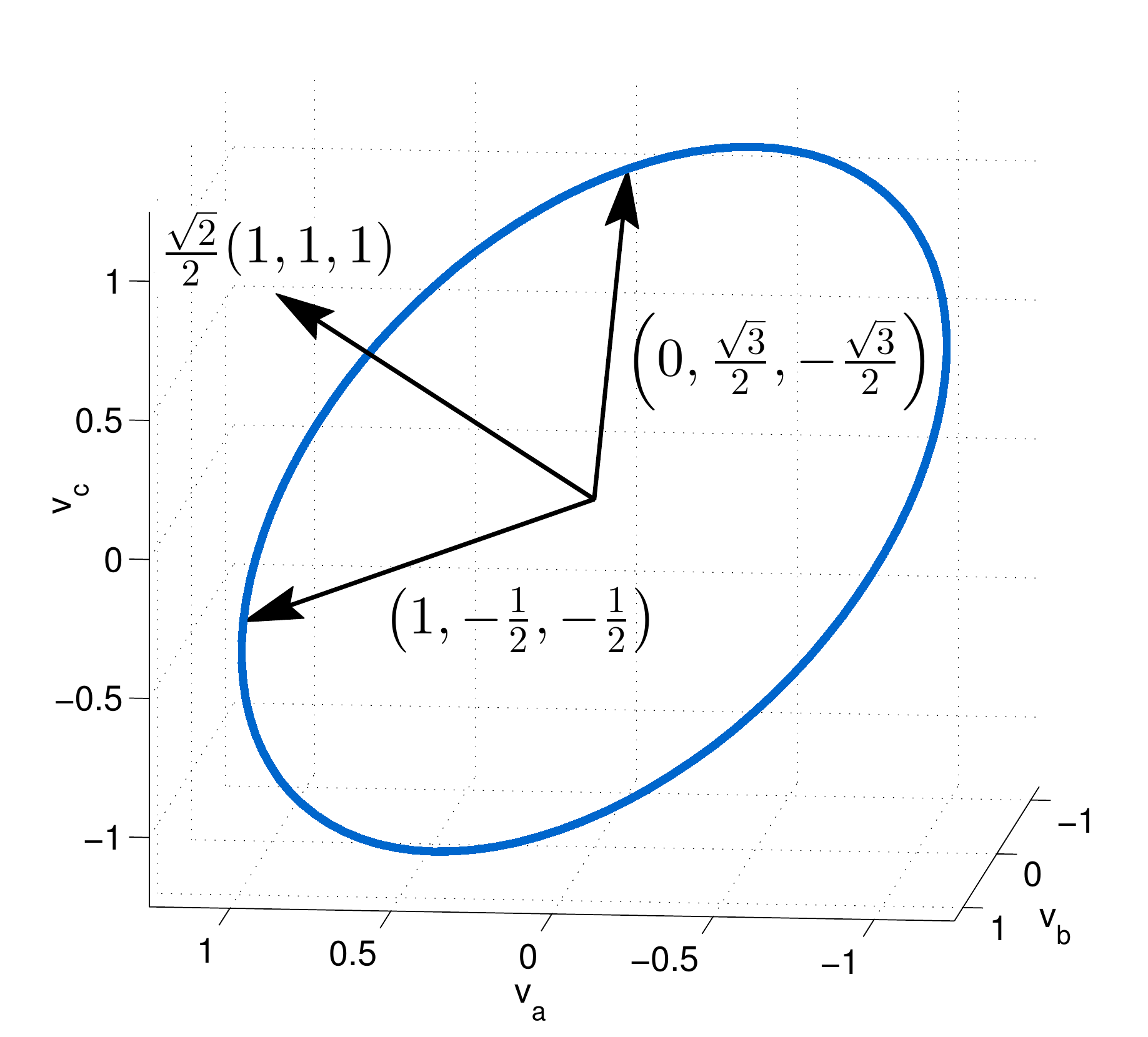}}
  \caption{A balanced three--phase power system with three principal axes. Notice that all  information is contained within a 2D subspace spanned by the eigenvectors $\big [1, \hspace{1mm} -\frac{1}{2},  \hspace{1mm}  - \frac{1}{2} \big]  $ and $\big [ 0, \hspace{1mm}  \frac{\sqrt 3}{2},  \hspace{1mm} -\frac{\sqrt 3}{2} \big] $, that is, in a space defined by the PCA which retains the two largest eigenvalues/eigenvectors of the corresponding covariance matrix, $\R_{\s}$. In other words, balanced three--phase power systems exhibit only two degrees of freedom.}
  \label{fig:PCA}
\end{figure}

Therefore, the remaining two eigenvectors of $\R_{\s}$ are $\q_2 = \v_r/ \| \v_r \|$ and $\q_3 = \v_i/ \| \v_i \|$ with the corresponding eigenvalues,  $\lambda_{2}= \frac{1}{4} \| \v_r\|$ and $\lambda_{3} = \frac{1}{4} \| \v_i\|$, so that the matrix of eigenvectors, $\Q^\T$, and the diagonal matrix of eigenvalues, $\mathbf{\Lambda}$, in \eqref{eq:R_s} take the form
\begin{align} \label{eq:Eigen}
\Q^\T = \sqrt{\frac{2}{3}}\begin{bmatrix}
\frac{\sqrt{2}}{2}& \frac{\sqrt{2}}{2} & \frac{\sqrt{2}}{2} \\
1 & \minus\frac{1}{2} & \minus\frac{1}{2}\\
0 & \frac{\sqrt{3}}{2} & \minus\frac{\sqrt{3}}{2}
\end{bmatrix} 
\hspace{12mm} 
\mathbf{\Lambda} = \frac{1}{4}\begin{bmatrix}
0& 0 & 0 \\
0 & 1.5 & 0\\
0 & 0 & 1.5
\end{bmatrix}.
\end{align}
Inspection of the diagonal elements of $\mathbf{\Lambda}$  in \eqref{eq:Eigen} reveals only two non-zero eigenvalues.  This verifies  Remark \ref{rem:rank} which states that the covariance  matrix of a three--phase power system voltage, $\R_{\s}$, is of Rank--2 and thus rank--deficient. The factor $\sqrt{2/3}$ which pre--multiplies $\Q^T$ in \eqref{eq:Eigen} serves to normalise the length of the eigenvectors to unity (ortho--normality).
\begin{remark}\label{rem:eigen}
 The matrix of eigenvectors,  $\Q^\T$ in \eqref{eq:Eigen}, is identical to the Clarke transformation matrix defined in \eqref{eq:clarke_1}.
Therefore,  all of the variance in three-phase power system voltages  can be explained by the two eigenvectors associated with the non-zero eigenvalues (principal axes) of the Clarke--transform--matrix. This offers the modern, Data Analytics, interpretation of Clarke's transform  as a Principal Component Analyser which performs a projection of three--phase power systems in  $\mathbb{R}^3$  onto a 2D subspace  spanned by the two largest orthogonal eigenvectors of the phase--voltage correlation matrix,  $[
1,  -\frac{1}{2} ,  -\frac{1}{2}]^\T $ and $[ 0,  \frac{\sqrt{3}}{2},  -\frac{\sqrt{3}}{2}]^T $,   as illustrated in  Figure \ref{fig:PCA}.
\end{remark}
\begin{remark} \label{Rem:Clarke_Park_Summary}
Remark \ref{rem:eigen} and Figure \ref{fig:PCA} offer a modern interpretation of the Clarke transform from a PCA--based dimensionality reduction viewpoint. Such new perspective opens numerous new avenues for the use of Data Analytics (such as Signal Processing and Machine Learning) in power grid research, and paves the way for innovative  transformation and analysis techniques for the future  Smart Grid -- not possible to achieve from the standard Circuit Theory principles.
\end{remark}
We shall next illuminate the power of Data Analytics in Smart Grid design and analysis,  by exploring new self--stabilising Clarke--inspired transforms, which unlike the original method adapt automatically to the dynamically unstable Smart Grid conditions. 
\section*{Meeting the Needs of Smart Grid: A Dynamic Clarke Transform for Unbalanced Power Systems} 
It is a pre--requisite for future smart grids to move away from the traditional high--inertia load--following operating strategy to a dynamic scenario which involves low inertia, smart loads and renewable generation, which causes the three-phase systems to operate in a dynamically unbalanced condition.  This, in turn, yields unequal phase voltages and non-uniform phase separations which need to be catered for in real time. \vspace{2mm}\\
\subsection*{Clarke and Symmetric transforms as a 3--point DFT} 
We shall now offer an interpretation of the the well--known inadequacy of current power system analysis techniques in unbalanced grid scenarios, through a link with the effects of incoherent sampling in spectral analysis.\\
\noindent {\bf Symmetrical Transform as a  Spatial Discrete Fourier Transform (DFT).}  The vector of three-phase time-domain voltages, $\s_k$ in \eqref{eq:3_phase},  is typically considered as a collection of three univariate signals. However,  observe that the phase voltage samples within $\s_k$ can also be treated as three samples of a  monocomponent signal rotating at a \emph{spatial frequency} $\Omega=-\frac{ 2 \pi}{3}$.   From this viewpoint,  the phasor vector, $\v$ in \eqref{eq:phasor_def_ori}, in a balanced system is given by 
\begin{align}
\label{Eq:Spatial_Sampling}
\v = \begin{bmatrix}
1, & e^{j\Omega},  & e^{j2\Omega} 
\end{bmatrix}^\T.
\end{align}
It is now obvious that $\v$ can be treated as a single sinusoid rotating at a spatial frequency of $\Omega = -\ph$, whereby the elements of $\v$ are  the corresponding phase voltages $v_{a,k},v_{b,k},$ and  $v_{c,k}$. \\
\begin{remark}\label{rem:first}
Under unbalanced conditions, the phasor vector, $\v$, does not represent a single complex-valued  spatial sinusoid since it contains the individual phasors with different  amplitudes and a non-uniform phase separation, as defined in \eqref{eq:phasor_def_ori}. \\
\end{remark}
%
%
%
Consider now the DFT of the phasor vector, $\v =[v_0, v_1, v_2]^\T \in \domC^{3 \times 1}$, given by 
\begin{align*}
X[k] = \frac{1}{\sqrt{3}}\sum_{n = 0}^{2} v_{n}e^{-j\frac{2\pi}{3}nk}, \quad k = 0, 1, 2 
\end{align*}
which can be expressed in an equivalent matrix form 
\begin{align}\label{eq:3pointDFT}
	\begin{bmatrix} 	X[0]  \\  X[1] \\ X[2] \end{bmatrix} {} & = {\frac{1}{\sqrt 3} \begin{bmatrix}   1   & 1 & 1 \\  1 & a & a^2 \\ 1 & a^2 & a \end{bmatrix}}
		  \begin{bmatrix} \bar V_{a} \\  \bar V_{b} \\ \bar V_{c} \end{bmatrix}
\end{align}
where $a = e^{-j\ph}$. The three-point DFT in \eqref{eq:3pointDFT}  therefore transforms the phasor vector $\v$  into a stationary component $X[0]$ and two other components, $X[1]$ and $X[2]$, which rotate at the respective spatial frequencies $\ph$ and $-\ph$. \\

\begin{remark}\label{rem:spatDFT}
 The spatial DFT in  \eqref{eq:3pointDFT} is  identical to the Symmetrical Component Transform in \eqref{eq:symmetrical_def}.  More specifically, the stationary DFT component, $X[0]$, corresponds to the  zero-sequence phasor, $\bar{V}_0$, while the fundamental DFT components, $X[1]$ and $X[2]$, represent respectively the positive-- and negative--sequence phasors.  This forms a basis for the treatment of three--phase component transforms from a Spectral Estimation perspective, and offers enhanced interpretation of the imperfections of these transforms in Smart Grid problems together with new avenues for the mitigation of these issues.\\
\end{remark}
\noindent {\bf Signal Processing interpretation.} Observe that the spatial sampling in \eqref{Eq:Spatial_Sampling} represents a crude critical sampling where the system frequency is  contained in the first component  of the underlying 3--point DFT, with no provision for the interpretation of drifting frequencies, as required by the Smart Grid. This explains the well known inability of the Symmetrical Component Transform to deal with transients in three--phase power systems, and the relation with incoherent sampling artefacts -- a standard issue in crudely sampled systems.\vspace{2mm}\\
\begin{tabular}{|p{0.95\columnwidth}|}
\rowcolor[HTML]{BBDAFF} \hline
\vspace{0.05cm} \bf Dealing with unbalanced phasors: The Symmetrical Transform \vspace{0.1cm}                \\
 \vspace{0.5mm}
The Symmetrical Transform was introduced by Charles Fortesque in 1918  to enable the decomposition of general unbalanced three-phase systems into three separate balanced networks  \cite{fortescue1918method}.  Unlike the Clarke and Park transforms, the symmetrical transform operates on the phasors (Fourier transforms) of the three-phase voltage, $\s_k$, and is given by 
%
%
%
%
\begin{equation}\label{eq:symmetrical_def}
	\left[\begin{array}{c} 	\bar V_0  \\  \bar V_{+} \\ \bar V_{-} \end{array}\right]  = {\frac{1}{\sqrt 3}  
	\underbrace{
	 \left[\begin{array}{c c c}   1   & 1 & 1 \\  1 & a & a^2 \\ 1 & a^2 & a \end{array} \right] }  
	 _{\mathrm{DFT~~matrix}}}
	 		 \left[\begin{array}{c} 
\bar{V}_a \\ \bar{V}_b \\  \bar{V}_c
\end{array}\right],  
\end{equation}
where $a = e^{-j\ph}$. The aim is to convert a general unbalanced phasor vector, $\v = \left[
\bar{V}_a, \: \bar{V}_b, \:  \bar{V}_c, \right]^\T$ in \eqref{eq:phasor_def_ori}, into three separate balanced components,  referred to as the zero--, positive-- and negative--sequence 
phasors, denoted respectively by $\bar V_0$, $\bar V_{+}$, and $\bar V_{-}$. Although the Symmetrical Component Transform can be used to analyse both balanced and unbalanced systems, it only applies to voltages in the phasor domain. 

\noindent {\bf Observe that the Clarke transform can be interpreted as the real part of the  3-point DFT matrix in \eqref{eq:symmetrical_def}, since the diagonalisation of the eigenvector matrix for circulant matrices yields the DFT matrix. }

Real-time Smart Grid tasks require analytical tools suitable for time-domain signals, thereby motivating the need for online dimensionality reduction techniques. \\ \hline 
\end{tabular}\\

\subsection*{A Data Analytics Interpretation}
We shall define the the imbalance ratios in unbalanced power systems as, $\delta_b = |\delta_b|e^{j \angle \delta_b}$ and $\delta_c= |\delta_c|e^{j\angle \delta_c}$. These ratios depend on the type of imbalance and yield a three--phase voltage covariance matrix
\begin{align} 
\label{Eq:Unbalanced_Covar_Matrix}
\R^u_{\s} 
%
= {} & \frac{1}{2}\begin{bmatrix}
1 & |\delta_b|\cos(\angle \delta_b) & |\delta_c| \cos(\angle \delta_c)  \\ 
 |\delta_b|\cos(\angle \delta_b) & |{\delta_b}|^2 & |\delta_b||\delta_c|\cos(\angle \delta_b - \angle \delta_c)  \\
 |\delta_c|\cos(\angle \delta_c) &  |\delta_b||\delta_c|\cos(\angle \delta_b - \angle \delta_c) &  |{\delta_c}|^2 \\
\end{bmatrix}
\end{align}
which is different from that for the balanced case in  \eqref{eq:R_def} and \eqref{eq:R_3}. \\

\begin{remark}\label{Rem:Unbalanced_Clarke}
Notice that for unbalanced power systems, due to the system imperfections modelled by the imbalance ratios $\delta_b = |\delta_b|e^{j \angle \delta_b}$ and $\delta_c= |\delta_c|e^{j \angle \delta_c}$, the eigenvector and eigenvalue matrices of the phase--voltage covariance matrix in \eqref{Eq:Unbalanced_Covar_Matrix} are  different from those  for the balanced system in \eqref{eq:R_def}.  This, in turn, implies that the projections within the Clarke matrix are no longer a perfect match for the three--phase voltages and also differ from the true Principal Components in data derived through PCA. 
\end{remark}

\begin{figure}[htbp]  \centering 
   {\includegraphics[clip = true, trim =0mm 0mm 0mm 0mm, scale=0.4]{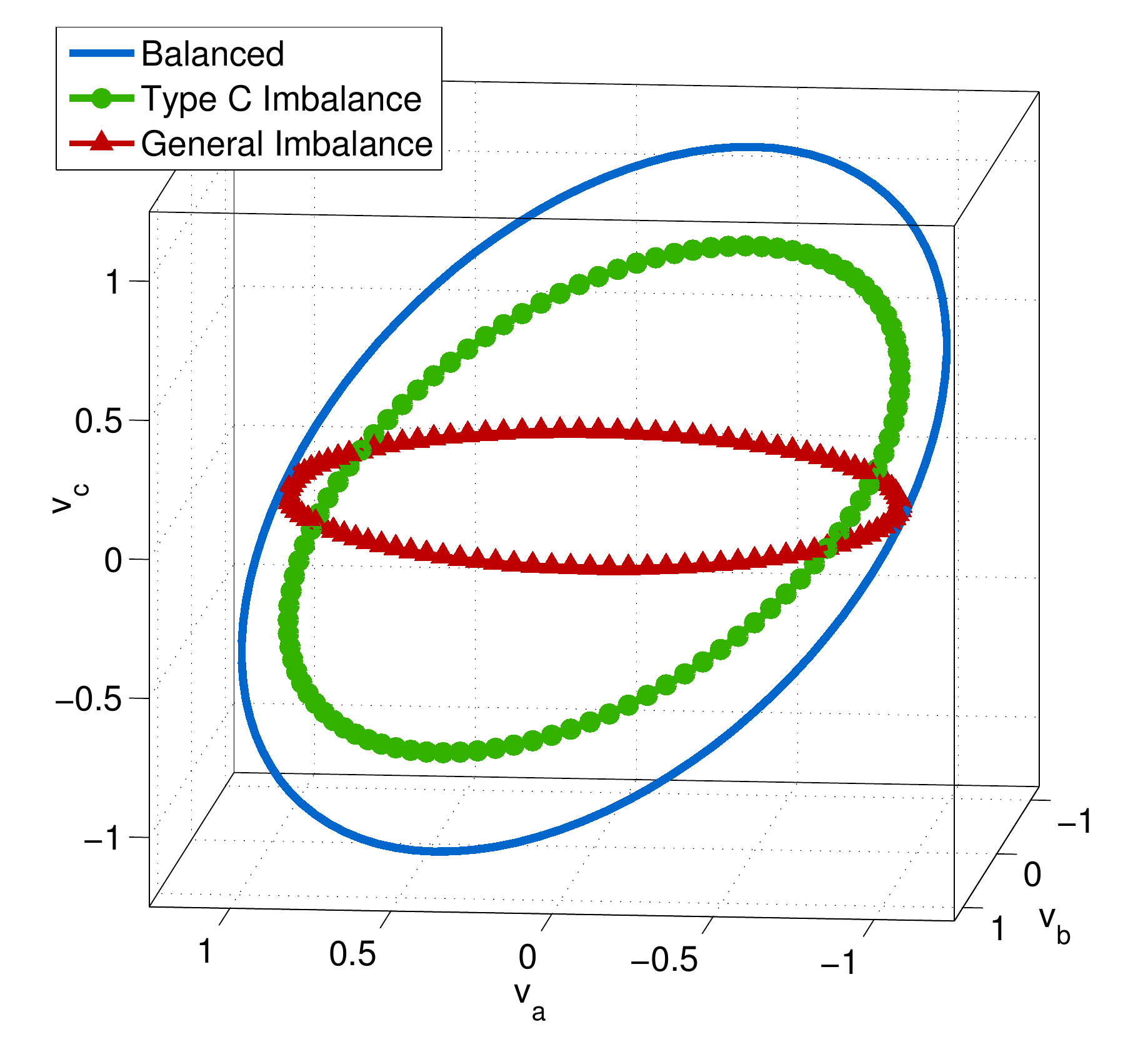}}
  \caption{Scatter plot trajectories  of the three-phase system voltages under: i) Balanced conditions (circle in blue); ii) Symmetric unbalanced condition for nominal frequency but unbalanced voltages (Type C voltage sag, ellipse in green); iii) General asymmetric imbalance for both off--nominal frequency and voltage imbalance (ellipse in red). Observe that independently of the type of imbalance, the Clarke voltages, $v_{\alpha}$ and $v_{\beta}$, will still reside in a 2D subspace of the 3D voltage space, but will no longer represent a perfect physically meaningful PCA--based dimensionality reduction scheme.  }
  \label{fig:genereal_R3}
\end{figure}

Figure \ref{fig:genereal_R3} illustrates that, regardless of the imbalance level in the  power system, the three-phase voltages still reside in a 2-dimensional subspace of $\mathbb{R}^3$. However, as the type and level of system unbalance dynamically change,  the ``static'' Clarke transform in \eqref{eq:clarke_red} will no longer be identical to the optimal ``correct'' PCA based dimensionality reduction scheme derived in  \eqref{eq:Eigen}.   This explains the well-known phenomenon that the application of the Clarke transform to unbalanced system voltages will spurious forms of  $\alpha\beta$  voltages \cite{Yili_GridSPM_2012,DPM_et_al:Frequency_Estimation}. 
\subsection*{Power System Imbalance  through the Lens of Complex Noncircular Statistics}
To further depict problems associated with unbalanced power systems, we shall revisit   the complex-domain representations of the Clarke and Park Transforms.

The Clarke's $\alpha\beta$ voltage in \eqref{eq:clarke_red} can be conveniently represented as a complex variable 
\begin{align} \label{eq:complex_clarke}
s_k \eqdef  v_{\alpha,k} + j v_{\beta,k}. 
\end{align}
or directly from the Clarke matrix
\begin{align} \label{eq:clarke_vec}
s_k  =   \c^\Her \s_k, \quad  \quad \c \eqdef \sqrt{\frac{2}{3}}\begin{bmatrix}
1, & e^{-j\ph}, &  e^{j\ph}\end{bmatrix}^\T
\end{align}
where $\c$ designates the complex Clarke transformation vector (see also \eqref{Eq:Spatial_Sampling}).  

Upon combining with the original phasors from \eqref{eq:3_phase_complex}, the complex $\alpha\beta$ voltage, $s_k$, assumes a physically meaningful representation through  the positive--sequence voltage, $\bar V_{+} $, and the negative--sequence voltage, $\bar V_{-} $, in the form 
\begin{equation} \label{eq:sequence_cmplx}
s_k  = \frac{1}{\sqrt{2}}\left(\bar V_{+} e^{j\omega k } + \bar V_{-}^*  e^{-j\omega k}\right),
\end{equation}
where  \cite{Paap_Symmetrical_2000}
\begin{align} \label{eq:posneg_def}
\bar V_{+} =  {} &  \frac{1}{\sqrt{3}}\left[V_{a}e^{j\phi_{a}}  + V_{b}e^{j\phi_{b}} + V_{c}e^{j \phi_{c}} \right]\\ \nonumber  
\bar V_{-}^*  = {}  &  \frac{1}{\sqrt{3}} \left[V_{a}e^{-  j\phi_{a}}  + V_{b}e^{-  j\left(\phi_{b} + \ph \right)} +  V_{c}e^{-  j\left( \phi_{c} - \ph \right)} \right]. 
\end{align}

\begin{remark}\label{Rem:Clarke_Pos_Neg_Sequence}
Notice that for balanced three--phase power systems, characterised by equal voltage magnitudes ($V_a = V_b = V_c$) and equal phase separations ($\phi_a = \phi_b = \phi_c$),  the negative sequence voltage sequence vanishes, that is,  $\bar V_{-} = 0$. This  yields the Clarke--transformed voltage for balanced power systems in the form
\begin{align} \label{eq:balanced_Clarke}
s_k = \bar V_{+} e^{j\omega k }. 
\end{align}
and a correct reading of the nominal system frequency. On the other hand, the unbalanced phase voltage conditions give rise to the negative sequence, $\bar V_{-}$, which results in a  bias in the estimation of the nominal system frequency, as the corresponding term $ e^{-j\omega k}$ in \eqref{eq:sequence_cmplx} rotates in the opposite direction of the true phasor, designated by $ e^{j\omega k}$.
\end{remark}

\begin{figure}[htbp]
\centering
  {\includegraphics[scale=0.4]{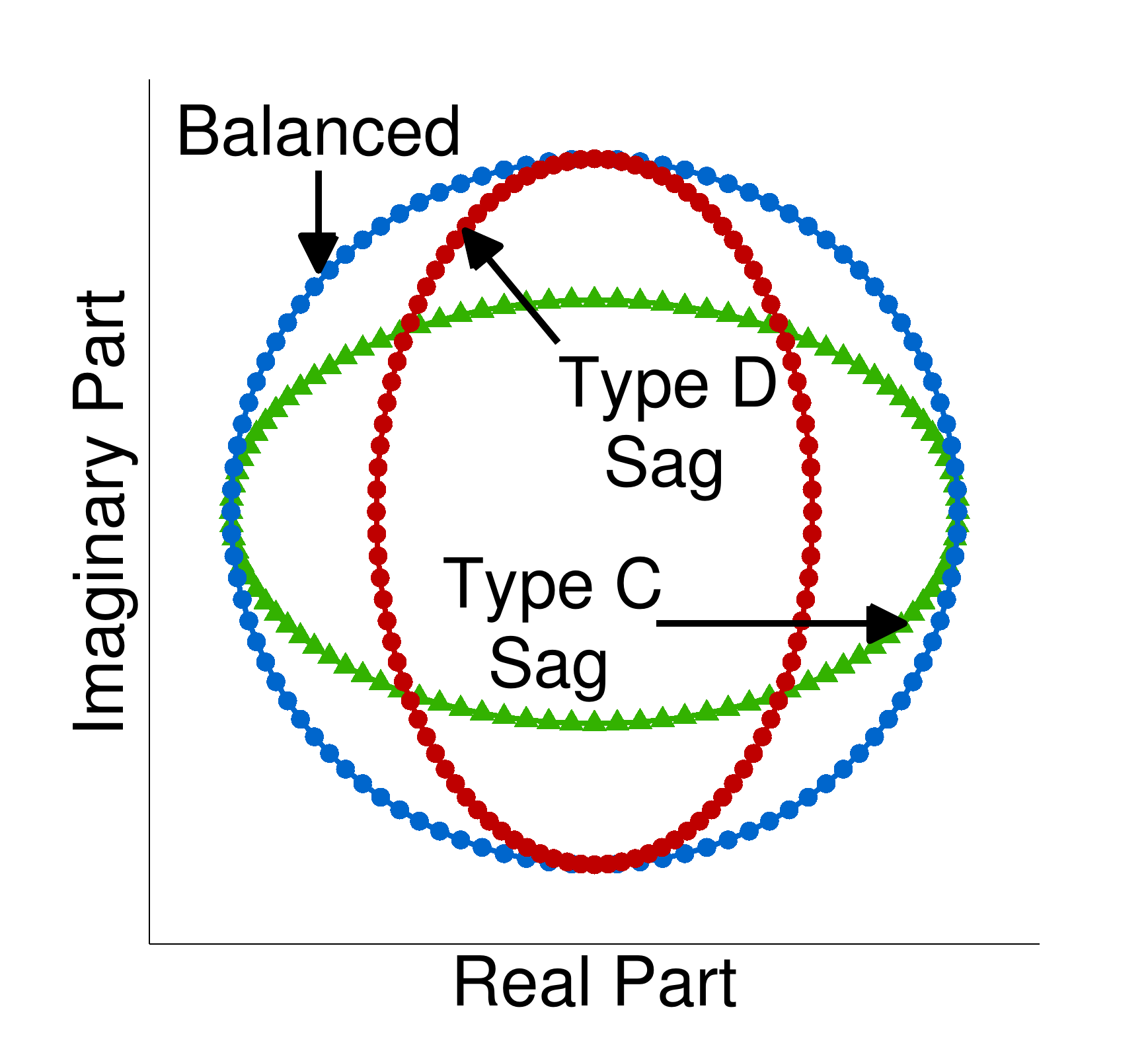}} \vspace{0cm}
  \caption{Scatter plot trajectories of the Clarke voltage $v_{k}= v_{\alpha}(k) + \jmath v_{\beta}(k)$. For a balanced system in \eqref{eq:3_phase}, which is characterised by the nominal frequency $\omega = \omega_{\circ}$, equal amplitudes of phase voltages $V_{a}=V_{b}=V_{c}$,  and equal phases  $\phi_{a} = \phi_{b} = \phi_{c}$, the trajectory of Clarke's voltage $v_{k}$ is circular (blue line). For unbalanced systems (in this case due to voltage sags), the Clarke voltage trajectories are noncircular (red and green ellipses). See Figure {\ref{fig:Phasor}} for more detail on voltage sags.}
  \label{fig:AB_trajectory}
\end{figure}

\noindent {\bf Complex Noncircularity as a Signature of Unbalanced Power Systems.} Figure \ref{fig:AB_trajectory} shows the scatter plot trajectories of the Clarke voltage in a balanced and two unbalanced system conditions. For a {\bf balanced power system}, the  scatter plot of  $s_k$ in \eqref{eq:balanced_Clarke} describes a circle, that is, it has {\bf only one degree of freedom}. In statistical terms, such random process  is called second--order circular (or proper) as it exhibits a rotation--independent distribution\footnote{The circularity diagram is a scatter plot of the real part versus the imaginary part of a complex variable. The strict definition of complex (non)circularity involves rotational invariance of the probability density function of a complex-valued random variable and is out of the scope of this article. For more detail, we refer the reader to \cite{Mandic_Book,schreier2010book}.}. 

\begin{figure}[ht]
\centering
  {\includegraphics[clip = true, trim= 25mm 20mm 10mm 20mm, scale=0.3]{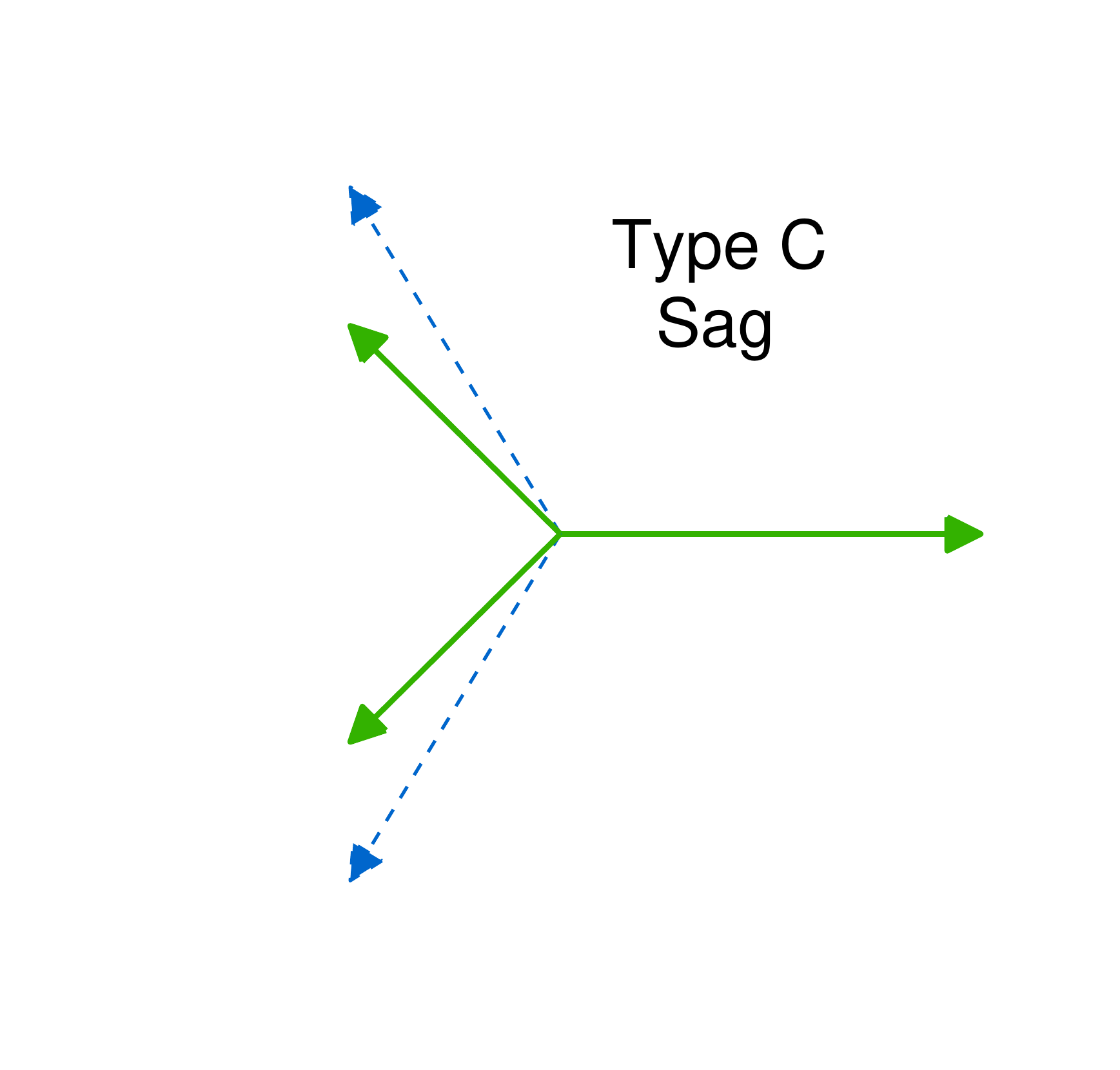}}\hspace{-5mm}
    {\includegraphics[clip = true, trim= 25mm 20mm 10mm 20mm, scale=0.36]{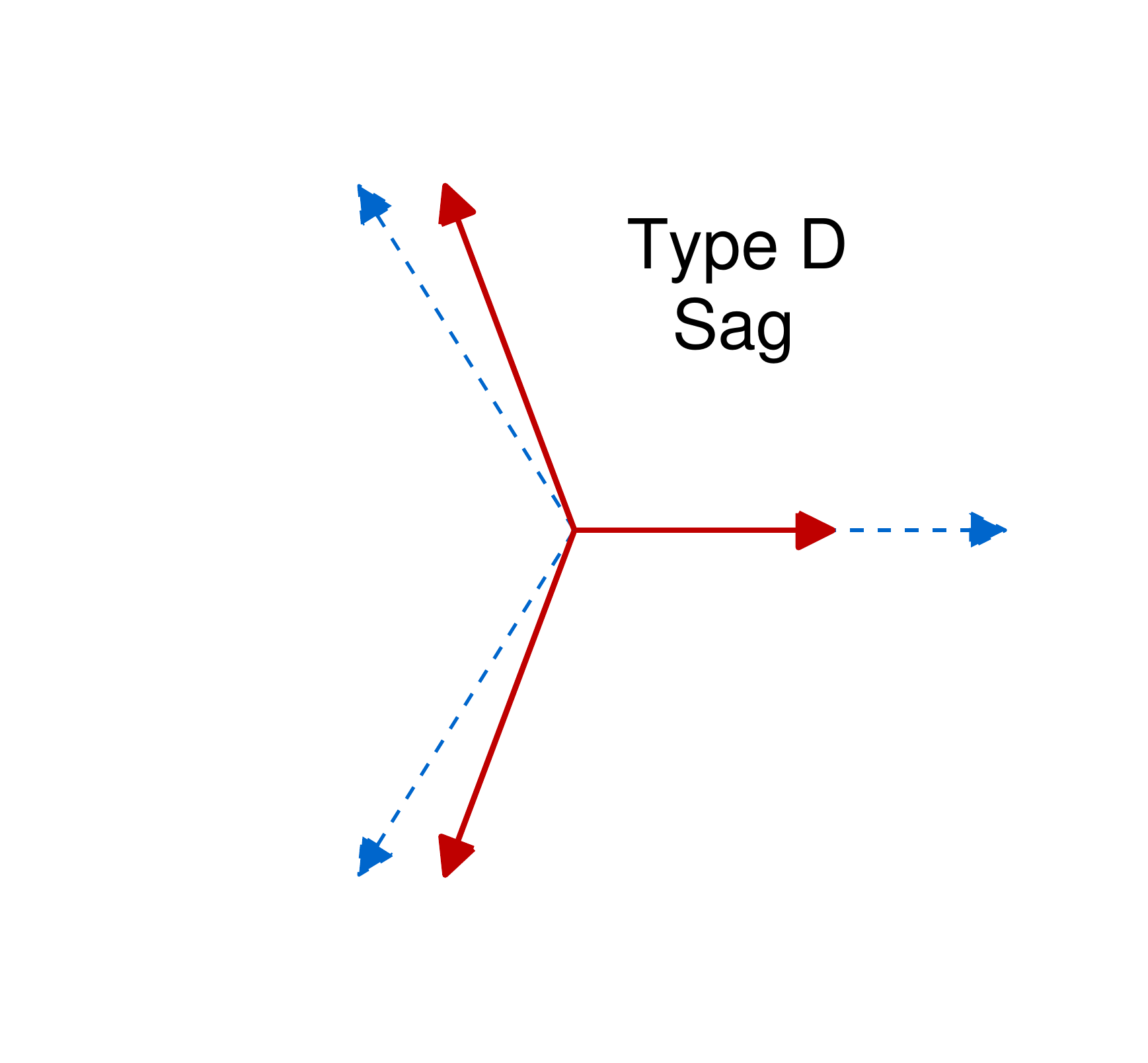}}
  \caption{Phasor diagrams of the effects of voltage sags. The dashed blue lines designate  a set of balanced  three-phase voltage phasors under nominal power system conditions, as in Figure {\ref{fig:frames}}. Notice the change in magnitudes and phase separations during  faults (Voltage sag C and Voltage sag D in this case).}
  \label{fig:Phasor}
\end{figure}

Recall from \eqref{eq:sequence_cmplx} that general {\bf dynamically unbalanced systems} are characterised by  $\bar V_{-}^* \ne 0$, that is, by {\bf two degrees of freedom} as exemplified by the  the two ellipses in Figure \ref{fig:AB_trajectory} which represent the trajectories for Type C and Type D voltage sags, well known power voltage imbalances further illustrated in the phasor diagram in Figure \ref{fig:Phasor}.  In statistical terms, this is reflected in $s_{k}$ assuming a rotation--dependent ``non--circular'' trajectory on the real--imaginary scatter diagram. This link with  noncircular complex statistics forms the basis for  simultaneous frequency estimation and fault detection in 3--phase unbalanced power systems {\cite{Bolen_Dip_Characterisation_2000, Bollen_Sags_1997,Bollen_SagsElsevier_2003,Sithan_TSIPN_2015, Xia_3Point_2014, Yili_GridSPM_2012, Xia_AMVDRGrid_2013,DPM_et_al:Frequency_Estimation}}, a key issue in modern low inertia power grids. 

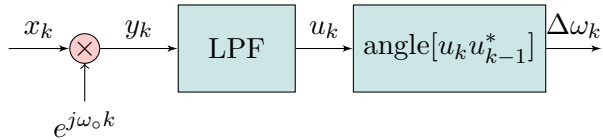
\begin{figure}[ht]
\centering
\tikzstyle{block} = [draw, fill=teal!20, rectangle, 
    minimum height=3em, minimum width=4em]
\tikzstyle{sum} = [draw, fill=red!20, circle, scale=1,inner sep=0pt, node distance=1cm]
\tikzstyle{input} = [coordinate]
\tikzstyle{output} = [coordinate]
\tikzstyle{pinstyle} = [pin edge={to-,thin,black}]

\begin{tikzpicture}[auto, node distance=2cm,>=latex']
    \node [input, name=input] {};
    \node [sum, right of=input, pin={[pinstyle]below:$e^{j\omega_{\circ}k}$}] (sum) {$\times$};
    \node [block, right of=sum] (controller) {$\text{LPF}$};
    \node [block, right of=controller, node distance=2.8cm] (system) {$\text{angle}[u_ku^*_{k-1}]$};
    \node [output, right of=system] (output) {};

    \draw [->] (controller) -- node[name=u] {$u_k$} (system);

    \draw [draw,->] (input) -- node {$x_k$} (sum);
    \draw [->] (sum) -- node {$y_k$} (controller);
    \draw [->] (system) -- node [name=y] {$\Delta \omega_k$}(output);
\end{tikzpicture}
\caption{Block diagram of a general fixed frequency demodulation scheme.}
\label{fig:Block_FM}
\end{figure}

\begin{remark}
Unbalanced system conditions introduce non--circular complex distributions which are characterised by two degrees  freedom, a scenario  for which conventional complex-valued linear estimation theory with only one available degree of freedom provides suboptimal solutions. Indeed, it was recently shown that the standard \emph{strictly linear} model when applied to the modelling of unbalanced systems in \eqref{eq:complex_clarke} is  inadequate, and a \emph{widely linear} model is required \cite{Picinbono_WL95, Yili_GridSPM_2012}. The notions of non--circularity and widely linear modelling underpin the proposed adaptive Clarke and Park transforms, explored in the next section. 
\end{remark}
\noindent {\bf Park Transform as an FM Demodulation Scheme.}
  Similar to the complex-valued representation of the Clarke Transform in \eqref{eq:complex_clarke},  the complex-valued version of the Park transform  in \eqref{eq:Park} is given by  
\begin{align}
v_k \eqdef {} & v_{d, k} + jv_{q, k}
\end{align}
which, in analogy to \eqref{eq:clarke_vec} can also be compactly represented as 
\begin{align} \label{eq:park_vec}
v_k = {} & e^{-j\omega_{\circ} k} \c^\Her \s_k = e^{-j\omega_{\circ} k} s_k,
\end{align} 
where $s_k = v_{\alpha, k} + j v_{\beta, k}$ is the Clarke voltage. Observe the ``circular'', time--varying, rotation frame designated by $e^{-j\omega_{\circ} k}$ which connects the Clarke and Park transforms.\\
\begin{remark}
 From a modern perspective viewpoint, the Park transform in  \eqref{eq:park_vec} can be interpreted as a frequency demodulation (FM) scheme \cite{Akke_FM_1997} of the $\alpha\beta$ voltage, whereby the demodulating frequency is the nominal system frequency $\omega_{\circ}$, as illustrated in Figure {\ref{fig:Block_FM}}. The demodulated instantaneous  frequency is then obtained  from the rate of change of the phase angles of the low-pass filtered signal $u_k$. \\
 \end{remark} 

%
%
%
%
%
%

Therefore, for a balanced three--phase power system operating at the fundamental frequency $\omega_{\circ}$, 
the Park transform yields the stationary positive sequence phasor, shown in Figure {\ref{fig:frames}} and given by  
\begin{align}
v_k = {} &  \bar V_{+}. 
\end{align}
{\bf Sources of bias in Park transform when used in  dynamically unbalanced Smart Grid.} In both current grids which incorporate renewables, and especially in the future Smart Grid, the three-phase voltages will be rarely perfectly balanced and the system frequency will never be at exactly the fundamental frequency \cite{Annette_unbalance2001}.  From \eqref{eq:sequence_cmplx}, the complex-valued $dq$ voltage  for a general unbalanced three-phase system,  which operates at an {\bf off-nominal} system frequency $\omega$, is given by
\begin{align}\label{eq:Park_out}
v_k = {} & \bar V_{+} e^{j(\omega - \omega_{\circ}) k } + \bar V_{-}^*  e^{-j(\omega + \omega_{\circ}) k}.
\end{align}   
 while the Park transform is designed for the nominal frequency, $\omega_{\circ}$.  Therefore,  the imperfect ``demodulation'' effect of the Park transform at an off-nominal frequency $\omega_{\circ} \neq \omega_{\circ}$ explains the spurious frequency estimation by the standard Park transform, as illustrated in Figure {\ref{fig:APT}}.\\
On the other hand, if an unbalanced system is operating at the {\bf nominal system frequency}, $\omega =\omega_{\circ}$, but {\bf off--nominal phase voltage/phase values}, the Park  $dq$ voltage in \eqref{eq:Park_out} becomes
\begin{align} \label{eq:Park_wSag}
v_k = {} & \bar V_{+} + \bar V_{-}^*  e^{-j2\omega_{\circ} k}. 
\end{align} 
which again is consistent with the output of an FM demodulator. This paves the way for the treatment of power system imbalances from the Communication Theory perspective, as summarised in Table \ref{tab:SigProc}. 
\begin{table}  \normalsize \centering
\caption{Signal processing interpretations of three-phase transformations.}
\bgroup
\def\arraystretch{1.2}
\begin{tabular}{| l | l |}\hline
 \textbf{Transform} & \textbf{Interpretation}  \\ \hline
 Symmetrical \cite{fortescue1918method}  & Spatial DFT  \\
Clarke \cite{Clarke_Book} & PCA \\
Park \cite{Park_1929} & FM demodulation \\ \hline
\end{tabular}
\egroup
\label{tab:SigProc}
\end{table}
\begin{remark} \label{Rem:Off_Nominal_Park}
Figure \ref{fig:unbalanced_frames} shows that during unbalanced system conditions, the optimal reference frames (basis vectors) for the $\alpha\beta$ and $dq$ voltages are different from the nominal ones defined by the classical Park and Clarke transforms.
\end{remark}

Table \ref{tab:comparison_old} summarises the functional expressions for the Clarke and Park transforms under both balanced and unbalanced conditions of the electricity grid.   As the Clarke and Park transforms will not yield accurate outputs under unbalanced system conditions (noncircular), their adaptive versions are required to enable: i)  accounting for voltage imbalances to provide a dynamically balanced Clarke output (circular), and ii) tracking instantaneous changes in system frequency through an ``adaptive"  Park transform. \\
\begin{table}[htbp]\centering
\bgroup
\normalsize
\def\arraystretch{1.2}
\begin{tabular}{|l |c |c|}\hline
 & \multicolumn{2}{c|}{\bf Power System Condition} \\\hline
 \bf Transform & \bf Balanced & \bf Unbalanced \\ \hline
Clarke \cite{Clarke_Book} & $\bar{V}_{+}e^{j \omega k }$ & $\bar{V}_{+}e^{j \omega k } + \bar{V}_{-}e^{-j \omega k } $ \\
Park \cite{Park_1929} & $\bar{V}_{+}$    &  $\bar V_{+} e^{j(\omega - \omega_{\circ}) k } + \bar V_{-}^*  e^{-j(\omega + \omega_{\circ}) k}$ \\ \hline
\end{tabular}
\egroup \vspace{1mm}
\caption{{Output of the Clarke and Park transformations.}}
\label{tab:comparison_old}
\end{table}
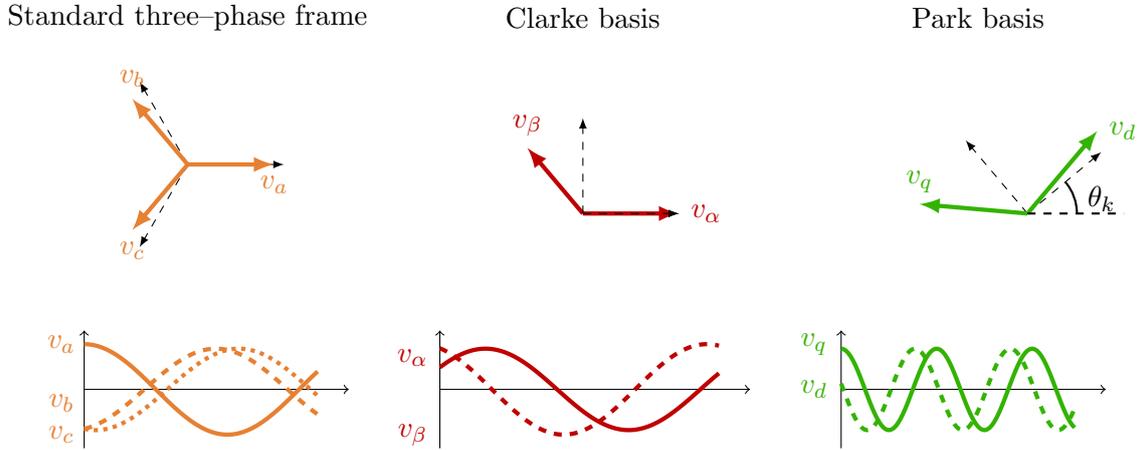
\begin{figure}[htbp]\centering
\definecolor{myblue}{rgb}{0, 0.4, 0.8}
\definecolor{myred}{rgb}{0.75, 0, 0}
\definecolor{mygreen}{rgb}{0.2, 0.7, 0}
\definecolor{myorange}{rgb}{0.9,0.5,0.2}
\begin{tikzpicture}[>=latex, xscale=0.65,yscale=0.65]
\draw[style=help lines] (0,0) (3,2);
\node[align=left] at (2,3) {Standard three--phase frame};
\node[align=left] at (10,3) {Clarke basis};
\node[align=left] at (18,3) {Park basis};
\coordinate (vec1) at (120:1.95); 
\coordinate (vec2) at (240:1.95);
\coordinate (vec3) at (0:1.95);
\coordinate (vec4) at (90:1.95);
\coordinate (vec5) at (40:1.95);
\coordinate (vec6) at (130:1.95);
\coordinate (vec7) at (175:2.2);
\coordinate (vec8) at (50:2.2);
\coordinate (vecunbal) at (130:1.75);
\coordinate (vecunbal2) at (230:1.75);
\coordinate (vecunbal3) at (0:1.75);
\draw[->,dashed,black] (2,0) -- ++(vec1) node[right] {};
\draw[->,dashed,black] (2,0) -- ++(vec2) node[below right] {};
\draw[->,dashed,black] (2,0) -- ++(vec3) node [below] {};

\draw[->,ultra thick,myorange] (2,0) -- ++(vecunbal2) node[below] {$v_c$};
\draw[->,ultra thick,myorange] (2,0) -- ++(vecunbal) node[above] {$v_b$};
\draw[->,ultra thick,myorange] (2,0) -- ++(vecunbal3) node [below] {$v_a$};

\draw[->, ultra thick, myred] (10,-1) -- ++(vec3) node[right] {$v_{\alpha}$};
\draw[->,ultra thick, myred] (10,-1) -- ++(vecunbal) node[above] {$v_{\beta}$};

\draw[->,dashed,black] (10,-1) -- ++(vec3) node[right] {};
\draw[->,dashed,black] (10,-1) -- ++(vec4) node[above] {};

\draw[->,dashed,black] (19,-1) -- ++(vec5) node[right] {};
\draw[->,dashed,black] (19,-1) -- ++(vec6) node[above] {};
\draw[->,ultra thick, mygreen] (19,-1) -- ++(vec8) node[right] {$v_d$};
\draw[->,ultra thick, mygreen] (19,-1) -- ++(vec7) node[above] {$v_q$};

\node[align=left] at (20.5,-0.7) {$\theta_k$};

\draw[thick,black, dashed] (19,-1) -- ++(vec3) node[] {};

\draw [thick] (20,-1) arc [start angle=0, end angle=40, radius=1]
    node [midway, right] {};
\end{tikzpicture}

\vspace{7mm}

\begin{tikzpicture}[xscale=0.6,yscale=0.6]

  \node[align=left, myred] at (7.2,0.7) {$v_{\alpha}$};
	\node[align=left, myred] at (7.2, -1) {$v_{\beta}$};

  \node[align=left, mygreen] at (16,1) {$v_q$};
\node[align=left, mygreen] at (16, 0) {$v_d$};

  \node[align=left, myorange] at (-0.5,1) {$v_{a}$};
    \node[align=left, myorange] at (-0.5, -0.3) {$v_{b}$};
  \node[align=left, myorange] at (-0.5, -1) {$v_{c}$};

  \draw[->] (0,-1.3) -- (0,1.3);
  \draw[->] (0,0)--(5.8,0);
  \draw[domain=0:5.12,samples=100,myorange, ultra thick] plot(\x,{cos(\x r)});
  \draw[domain=0:5.12,samples=100,myorange, ultra thick, dashed] plot(\x,{0.9*cos((\x-2.89) r)});
  \draw[domain=0:5.12,samples=100,myorange, ultra thick, dotted] plot(\x,{0.9*cos((\x +2.89) r)});
  
    \draw[->] (7.8,-1.3) -- (7.8,1.3);
  \draw[->] (7.8,0)--(14.6,0);
  \draw[domain=7.8:14.92-1,samples=100,myred, ultra thick] plot(\x,{0.9*cos((\x-8.8) r)});
  \draw[domain=7.8:14.92-1,samples=100,myred, ultra thick, dashed] plot(\x,{sin((\x-5.8) r)});

      \draw[->] (16.6,-1.3) -- (16.6,1.3);
  \draw[->] (16.6,0)--(22.4,0);
    \draw[domain=16.6:14.92-2+8.8,samples=100,mygreen, ultra thick] plot(\x,{0.9*cos(3*(\x-16.6) r)});
      \draw[domain=16.6:14.92-2+8.8,samples=100,mygreen, ultra thick, dashed] plot(\x,{0.9*sin(3*(\x-15.6) r)});

\end{tikzpicture}
\caption{Effects of unbalanced three--phase power systems on the accuracy of classical three-phase reference frames for their analysis. Observe that both standard Clarke and Park reference frames are unsuitable for  unbalanced phase voltages/phases and the operation at off-nominal frequencies, as exemplified by the oscillatory Park output for a static off--nominal frequency, instead of the two straight lines, as in Figure {\ref{fig:frames}}.}
\label{fig:unbalanced_frames}
\end{figure}
\section*{Teaching Old Power Systems New Tricks: Adaptive Clark \& Park Transforms}
It has been widely accepted that the minimum mean square error (MMSE) estimator for a complex--valued process, $y_{k}$, based on a complex--valued regressor, ${\bf x}_{k}$ is a straightforward extension of the corresponding real--valued one, and assumes the {\bf strictly linear} form
\begin{equation}
\label{Eq:SL_Estimator}
{\hat y}_{k}= E\{ y_{k} | {\bf x}_{k} \} =  {\bf h}^{H} {\bf x}_{k}	
\end{equation}
where ${\bf h}$ is a set of complex--valued coefficients. However, given that 
\begin{equation}
	{\hat y}_{r,k}= E\{y_{r,k}| {\bf x}_{r,k} ,{\bf x}_{i,k} \} \quad {\hat y}_{i,k}= E\{y_{i,k}| {\bf x}_{r,k} , {\bf x}_{i,k} \}
\end{equation}
and using the well known identities, $x_{r}= (x + x^{*})/2$ and $x_{i}= (x - x^{*})/2 \jmath$, the correct estimator for the generality of complex data is {\bf widely linear}, and is given by {\cite{Picinbono94,Mandic_Book}}
\begin{equation}
\label{Eq:WL_Estimator}
{\hat y}_{k} = {\bf h}^{H} {\bf x}_{k} + {\bf g}^{H} {\bf x}_{k}^{H}	
\end{equation}
A  comparison with unbalanced power systems shows  that the general unbalanced $\alpha \beta$ voltage in \eqref{eq:sequence_cmplx}, which is a sum of two complex-valued sinusoids rotating in opposite directions,  can be represented by a widely linear autoregressive (WLAR) model, given by {\cite{Picinbono94,Mandic_Book,Yili_GridSPM_2012}}
\begin{align} \label{eq:WLAR} 
	{s}_k = {} &  h^*s_{k-1} + g^*s^*_{k-1} ,
\end{align}
\begin{remark} \label{Rem:WL_Freq}
A comparison with \eqref{eq:sequence_cmplx} and \eqref{eq:Park_out}  shows that the WLAR coefficients $h, g \in {\mathbb C}$ contain the information related to the system frequency, $\omega$, together with providing the desired additional degree of freedom -- a pre--requisite for the analysis of unbalanced power system.
\end{remark}
The level of voltage imbalance in the power system can be defined through the voltage unbalance factor (VUF), given by \cite{Annette_unbalance2001}
\begin{align} \label{eq:VUF_def}
\mathrm{VUF:} \qquad \kappa \eqdef {} & \bar{V}_{-}/  \bar{V}_{+}.
\end{align}
so that, as showed in \cite{Yili_GridSPM_2012}, the system frequency can be expressed through the WLAR coefficients, $h$ and $g$, and the VUF, $\kappa$, to yield
\begin{align}\label{eq:simul_WLAR}
e^{j\omega} = h^* + g^*\kappa   \quad \text{and} \quad e^{-j\omega} = h^* + \frac{g^*}{\kappa^*}.
\end{align}
It is snow straightforward to solve for   the  system frequency, $\omega$, and VUF as 
\begin{align} \label{eq:omega_WLAR} 
e^{j\omega} = {} & \Real{h} + j\sqrt{ \mathsf{Im}^2\{h\} - |g|^2 },  \\\label{eq:kappa_WLAR}
\kappa = \frac{\bar V_{-}}{\bar V_{+}} = {} & \frac{j}{g^*}\left(\Imag{h} + \sqrt{ \mathsf{Im}^2\{h\} - |g|^2 }\right). 
\end{align}
\subsection*{Self--Balancing Clarke and Park Transforms}
The knowledge of the VUF, $\kappa$ in \eqref{eq:kappa_WLAR}, proves instrumental in eliminating the negative sequence phasor, $\bar{V}_{-}$, from the $\alpha\beta$ voltage $s_k$. To this end, consider the expression \cite{Xia_Balancing_2014}
\begin{align} \label{eq:balanced_m}
m_k \eqdef {} & \sqrt{2}\left( s_k - \kappa^*s^*_k \right)\\ \nonumber
    = {} & \bar V_{+} e^{j\omega k } + \bar V_{-}^*  e^{-j\omega k} - \frac{\bar V^*_{-}}{\bar V^*_{+}}\left(\bar V^*_{+} e^{-j\omega k } + \bar V_{-}  e^{j\omega k}  \right) \\
\label{eq:balanced_m2}   = {} & \bar V_{+}\left(1- |\kappa|^2   \right)e^{j\omega k} . 
\end{align}
whereby the value of $\kappa$ is readily available from the WLAR coefficients  in \eqref{eq:kappa_WLAR}. This makes it possible to eliminate the effects of voltage imbalance on the Clarke's $\alpha \beta$ voltage  in the form
\begin{align}\label{eq:balanced_m3}
\bar{m}_k =  m_k/(1- |\kappa|^2 ) = \bar V_{+}e^{j\omega k}. 
\end{align}
\begin{remark}
 The voltage $\bar{m}_k$ contains only the positive voltage sequence $\bar V_{+}$ and is immune to the effects of system imbalance, reflected  through the non--zero negative sequence $\bar V_{-}$. The operation in \eqref{eq:balanced_m3} can be  regarded as an adaptive Clarke transform, the output of which is always identical to the correct $\alpha\beta$ voltage for a balanced system.    
\end{remark}


Finally,  from the estimated time-varying values of the drifting system frequency (through  $e^{j\omega_k}$ in \eqref{eq:omega_WLAR}  and $\kappa_k$ in  \eqref{eq:kappa_WLAR}), the adaptive Clarke and Park  transforms can be summarised as  {\cite{Kanna_Stabilising_Clarke_El_Letters_2017}}
\begin{subequations} 
\begin{align}  \label{eq:adaptive_Clarke}
\mathrm{Adaptive~Clarke~transform: } \quad \bar{m}_k = {} & \sqrt{2}(s_k - \kappa_k^* s^*_k)/( 1 - |\kappa_k|^2) \\ \label{eq:adaptive_Park}
\mathrm{Adaptive~Park~transform: }  \quad   \tilde{m}_k = {} & e^{-j\omega_k k}\bar{m}_k ,
\end{align}
\end{subequations} 
where $\bar{m}_k$ is the adaptive $\alpha\beta$ (Clarke) voltage  while $\tilde{m}_k$ is the adaptive $dq$ (Park)  voltage.\\
For real--time adaptive operation, the adaptive  Clarke and Park transforms can  be implemented based on \eqref{eq:adaptive_Clarke}--\eqref{eq:adaptive_Park}, and using a suitable adaptive learning algorithm (e.g. least mean square (LMS) or Kalman filter) to track the VUF, $\kappa_k$, and system frequency, $\omega_k$.  For illustration, we present the adaptive Clarke/Park transform in Algorithm \ref{alg:adaptivePark}, with the  augmented complex least mean square (ACLMS) {\cite{Javidi_ACLMS_08,Mandic_Book}} used to estimate the information bearing WLAR coefficients $h$ and $g$. 
\begin{remark}
The adaptive ``balancing'' versions of the Clarke and  Park transforms  perform accurately the  respective dimensionality reduction and rotation operations, regardless of the drifts in system frequency or level of voltage/phase imbalance. Table \ref{tab:comparison} summarises the functional expressions for these adaptive transforms, which  make it possible to  use standard analysis techniques designed for nominal conditions in general unbalanced systems, resulting in a bias-free operation. 
\end{remark}
\begin{table}[htbp]\centering
\bgroup
\normalsize
\def\arraystretch{1.2}
\begin{tabular}{|l |c |c|}\hline
 & \multicolumn{2}{c|}{\bf Power System Condition} \\\hline
 \bf Transform & \bf Balanced & \bf Unbalanced \\ \hline
Clarke \cite{Clarke_Book} & $\bar{V}_{+}e^{j \omega k }$ & $\bar{V}_{+}e^{j \omega k } + \bar{V}_{-}e^{-j \omega k } $ \\
Balancing \cite{Xia_Balancing_2014} & $\bar{V}_{+}e^{j \omega k }$ & $(1 - |\kappa|^2)\bar{V}_{+}e^{j \omega k } $ \\
\emph{Adaptive Clarke}  & $\bar{V}_{+}e^{j \omega k } $ & $\bar{V}_{+}e^{j \omega k } $ \\ \hline
Park \cite{Park_1929} & $\bar{V}_{+}$    &  $\bar V_{+} e^{j(\omega - \omega_{\circ}) k } + \bar V_{-}^*  e^{-j(\omega + \omega_{\circ}) k}$ \\ 
\emph{Adaptive Park}  & $\bar{V}_{+}$ & $\bar{V}_{+} $ \\ \hline
\end{tabular}
\egroup \vspace{1mm}
\caption{{Comparison of the classical static three-phase transforms.}}
\label{tab:comparison}
\end{table}

\begin{figure}\centering
\includegraphics[scale=0.55, clip = true, trim = 0mm 0mm 0mm 8mm]{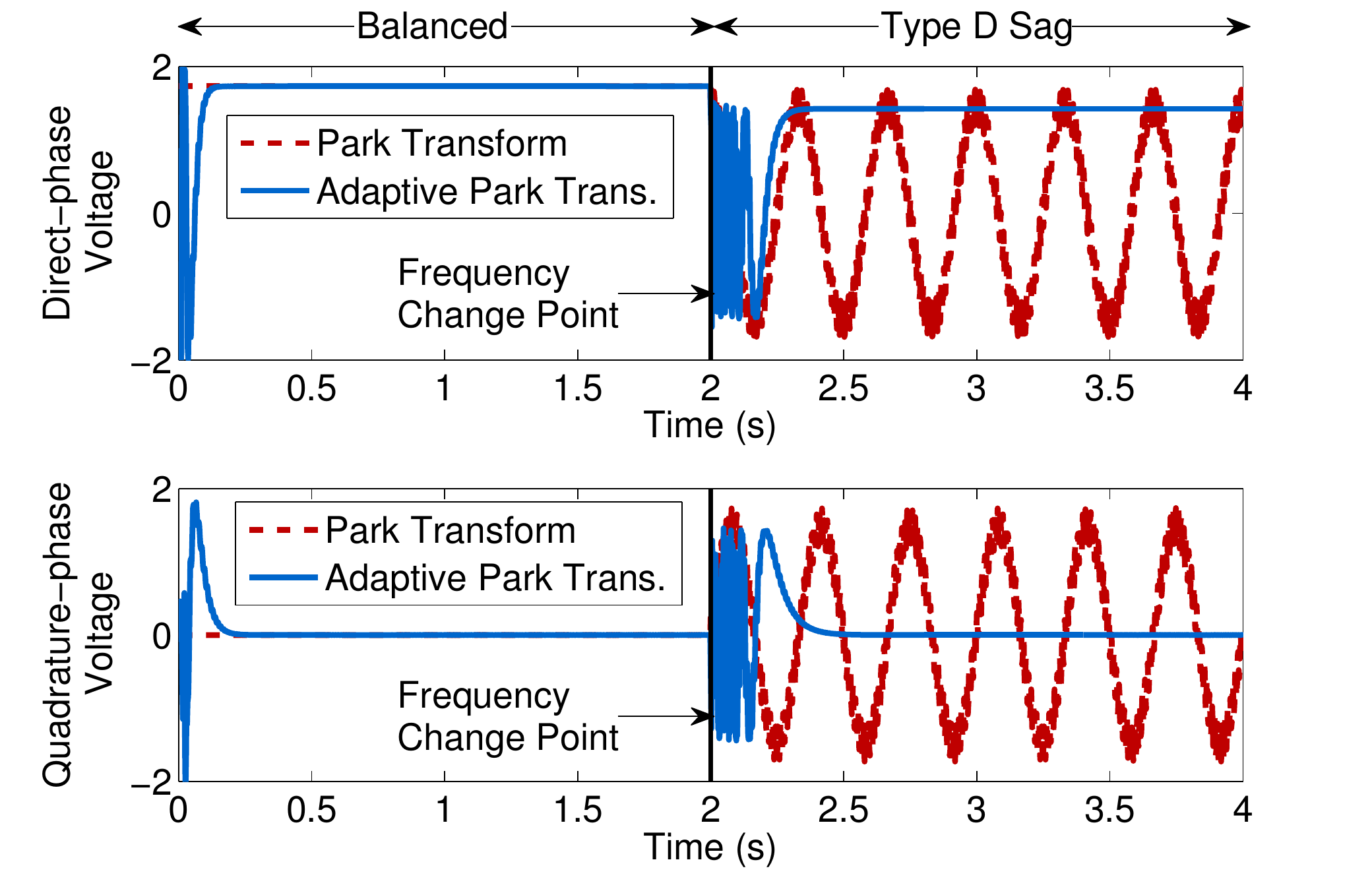}
\caption{Operation of the classical and adaptive Park Transforms in frequency estimation in unbalanced power grids. Observe the self--stabilising nature of the adaptive Park transform in the presence of system imbalance (frequency drop) starting from $t=2s$.}
\label{fig:APT}
\end{figure}

\begin{figure}[htbp]\centering  
    {\includegraphics[clip = true, trim =5mm 100mm 10mm 0mm, scale=0.55]{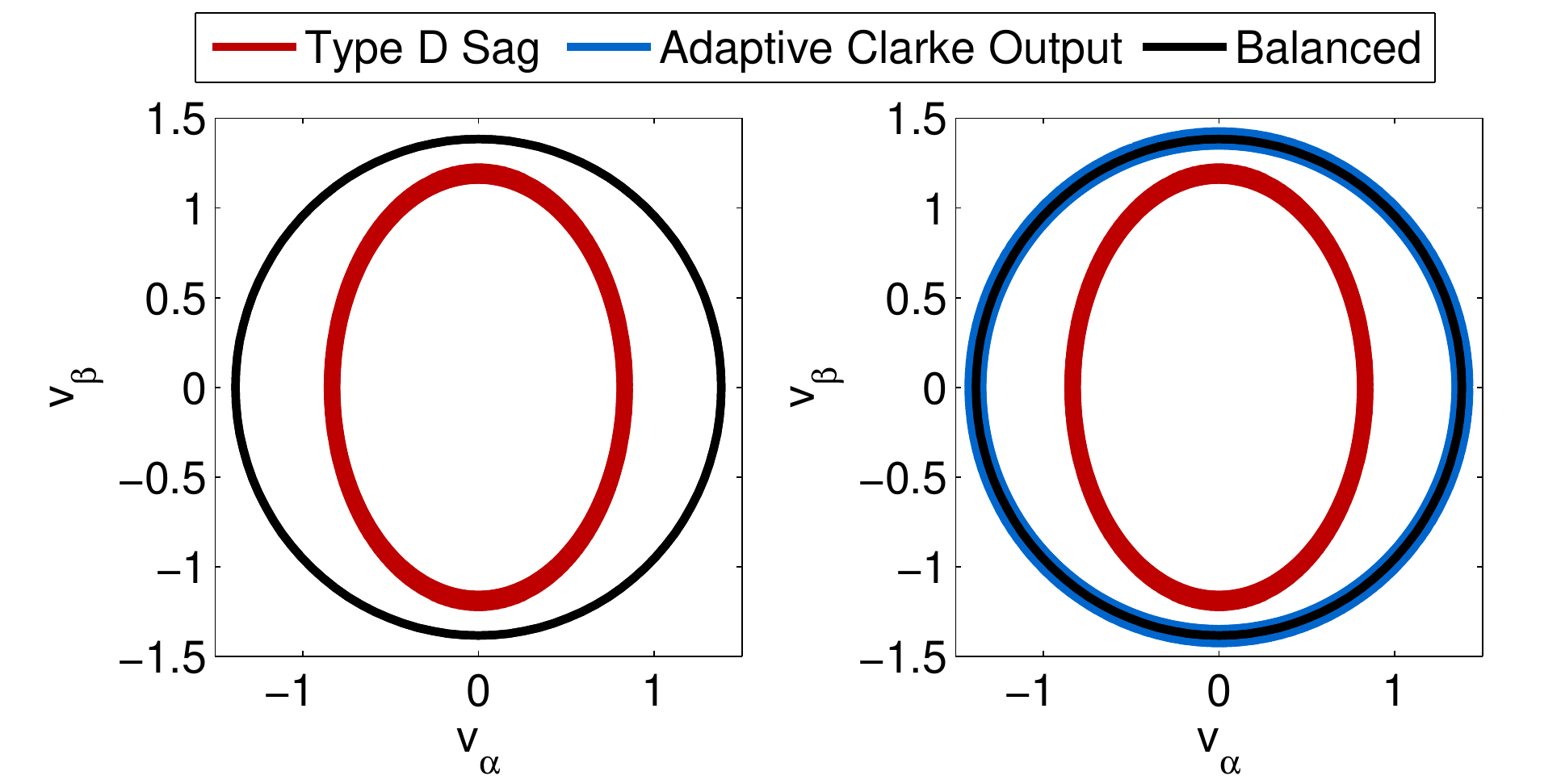}}\vspace*{-3mm}
  \caption{Self--balancing nature of the adaptive Clarke transform at a nominal system frequency, $\omega_{\circ}$, but in the presence of Type D voltage sag, indicated by the red ellipse. The adaptive Clarke transform yields a circular (balanced) $\alpha\beta$ voltage, ${\bar m}_{k}$ in \eqref{eq:adaptive_Clarke} (blue circle), which coincides with the optimal ``balanced'' conditions (black circle).}
  \label{fig:Circ_park}
\end{figure}

\begin{algorithm}[htbp] 
 \caption{\hspace*{-1mm}{\bf.}   Adaptive Clarke/Park Transform}
\textbf{Input}: Original three-phase voltages, $\s_k$, learning rate, $\mu$ \\
\textbf{At each time instant} $k > 0$ : 
\begin{algorithmic}[1]
\State 	Obtain the Clarke transform : $ s_k = \sqrt{2}\,\c^\Her\s_k  $
\State 	Update the weights of ACLMS   
\begin{itemize}
\item[] $\begin{array}{l c l}
\varepsilon_k & = {} & s_k    - (h^*_{k-1}s_{k-1} +  g^*_{k-1}s^*_{k-1}) \\ 
h_{k} &= {} &  h_{k-1} + \mu s_{k-1} \varepsilon^*_k \\ 
g_{k} &= {} &  g_{k-1} + \mu s^*_{k-1} \varepsilon^*_k \\
\end{array}
$
\end{itemize}
\State Use \eqref{eq:adaptive_Clarke} and \eqref{eq:adaptive_Park} to obtain $\kappa_k$ ad $e^{j{\omega}_k}$ 
\State 	Calculate adaptive Clarke transform: $\bar{m}_k =  (s_k - \kappa^*_k s^*_k) /( 1 - |\kappa_k|^2)$

\State 	Calculate adaptive Park transform: $\tilde{m}_k = {}  e^{-j{\omega_k}k}\bar{m}_k$

\end{algorithmic} 
\label{alg:adaptivePark}
\end{algorithm}
Figure \ref{fig:APT} shows the direct and quadrature Park voltages, $v_{d, k}$ and $v_{q, k}$, obtained from both the original Park transform defined in  ({\ref{eq:Park}}), and the adaptive Park transform, $\tilde{m}_k$  in \eqref{eq:adaptive_Park} and Algorithm 1. Observe the oscillating output of the original Park transform (broken red line) when the system frequency suddenly changed to a lower,  off-nominal, value  starting from $t=2s$. On the other hand,  the adaptive Park transform was able to converge to a stationary (non-oscillatory) correct phasor soon after this system imbalance.
Figure \ref{fig:Circ_park} further supports the ``balancing" nature of the Adaptive Clarke transform through the corresponding circularity diagrams of $v_{\alpha}$ on the $x$-axis versus $v_{\beta}$ on the $y$-axis. The circular profile for the adaptive $\alpha\beta$ voltage, ${\bar m}_{k}$ in \eqref{eq:adaptive_Clarke},  in the presence of Type D voltage sag indicates its successful operation.
\section*{Conclusion}
The operation of the future and almost permanently dynamically unbalanced Smart Grids requires close cooperation and convergence between the Power Systems and Data Analytics communities, such as those working in Signal Processing and Machine Learning.  A major prohibitive factor in this endeavour has been a lack of common language;  for example, the most fundamental techniques, such  the Clarke and Park transform introduced respectively in 1943 and 1929, have been designed from a Circuit Theory perspective and only for balanced ``nominal'' system conditions, characterised by high grid inertia. This renders such methodologies both inadequate for the demands of modern, dynamically unbalanced, Smart Grids and awkward for linking up with Data Analytics communities. To help bridge this gap, we have provided modern interpretations of the the Clarke and related transforms through the modern subspace, demodulation, and complex non--circularity concepts. These have served  as a mathematical lens into the inadequacies of current methodologies under unbalanced power system conditions, and have enabled us to create a framework for the understanding and mitigation of the effects of off--nominal system frequency and  dynamically unbalanced phase voltages and phases.  
All in all, such a conceptual insight permits seamless migration of ideas, in a bidirectional way, between these normally disparate communities and helps demystify power system analysis for Data Analytics practitioners.  

It is fitting to conclude the article with a quote from J.E. Brittain's article on Edith Clarke \cite{Brittain_Clarke_85}, which states 
\begin{quote}
 \emph{She [Clarke] translated what many engineers found to be esoteric mathematical methods into graphs or simpler forms during a time when power systems were becoming more complex and when the initial efforts were being made to develop electromechanical aids to problem solving.}
\end{quote}
 It is our hope that this modern perspective of the Clarke and related transforms will help extend the legacy of Edith Clarke well into the Information Age, in addition to empowering  analysts with enhanced intuition and freedom in algorithmic design. It further opens up new possibilities in the otherwise prohibitive applications of Clarke--inspired transforms in future low inertia Smart Grids.

 \balance
\bibliographystyle{ieeetr}
\bibliography{sithan_ref}

\begin{thebibliography}{10}

\bibitem{Brittain_Clarke_85}
J.~E. Brittain, ``{From Computor to Electrical Engineer: The Remarkable Career
  of Edith Clarke},'' {\em IEEE Transactions on Education}, vol.~28,
  pp.~184--189, Nov 1985.

\bibitem{E_Clarke:Human_Computer}
E.~Clarke, ``Calculator,'' {US P}atent 1552113, September 1925.

\bibitem{Cantelli_ClarkePID_06}
M.~Canteli, A.~Fernandez, L.~Eguiluz, and C.~Estebanez, ``Three-phase adaptive
  frequency measurement based on {C}larke's transformation,'' {\em IEEE
  Transactions on Power Delivery}, vol.~21, no.~3, pp.~1101--1105, 2006.

\bibitem{Barbosa_FuzzyClarke_2011}
D.~Barbosa, U.~C. Netto, D.~V. Coury, and M.~Oleskovicz, ``Power transformer
  differential protection based on clarke's transform and fuzzy systems,'' {\em
  IEEE Transactions on Power Delivery}, vol.~26, pp.~1212--1220, April 2011.

\bibitem{Yili_GridSPM_2012}
Y.~Xia, S.~Douglas, and D.~Mandic, ``Adaptive frequency estimation in smart
  grid applications: Exploiting noncircularity and widely linear adaptive
  estimators,'' {\em IEEE Signal Processing Magazine}, vol.~29, no.~5,
  pp.~44--54, 2012.

\bibitem{Arnold_Proc_IEEE_2011}
G.~W. Arnold, ``Challenges and opportunities in smart grid: A position
  article,'' {\em Proceedings of the IEEE}, vol.~99, no.~6, pp.~922--927, 2011.

\bibitem{Bollen_2001}
M.~H.~J. Bollen, ``Voltage sags in three-phase systems,'' {\em IEEE Power
  Engineering Review}, vol.~21, no.~9, pp.~8--15, 2001.

\bibitem{Dora_Chat_2011}
D.~Nakafuji, ``Personal communication,'' heco, hi, usa, 2011.

\bibitem{Bollen_DSP_Power_IEEE_SPM_2009}
M.~H.~J. Bollen, I.~Y.~H. Gu, S.~Santoso, M.~F. McGranaghan, P.~A. Crossley,
  M.~V. Ribeiro, and P.~F. Ribeiro, ``Bridging the gap between signal and
  power,'' {\em IEEE Signal Processing Magazine}, vol.~26, no.~4, pp.~11--31,
  2009.

\bibitem{DPM_et_al:Frequency_Estimation}
D.~P. Mandic, Y.~Xia, and D.~Dini, ``Frequency estimation,'' {US P}atent
  9995774, June 2018.

\bibitem{Tirza_Imbalance_Symchorphasor_IEEE_TPS_2018}
T.~Routtenberg and Y.~C. Eldar, ``Centralized identification of imbalances in
  power networks with synchrophasor data,'' {\em IEEE Transactions on Power
  Systems}, vol.~33, pp.~1981--1992, March 2018.

\bibitem{Routtenberg_KCL_2013}
T.~Routtenberg and L.~Tong, ``Joint frequency and phasor estimation under the
  {KCL} constraint,'' {\em IEEE Signal Processing Letters}, vol.~20,
  pp.~575--578, June 2013.

\bibitem{Pradhan_LMS_2005}
A.~Pradhan, A.~Routray, and A.~Basak, ``Power system frequency estimation using
  least mean square technique,'' {\em IEEE Transactions on Power Delivery},
  vol.~20, no.~3, pp.~1812--1816, 2005.

\bibitem{Routray_EKFSLAR2_2002}
A.~Routray, A.~K. Pradhan, and K.~P. Rao, ``A novel {K}alman filter for
  frequency estimation of distorted signals in power systems,'' {\em IEEE
  Transactions on Instrumentation and Measurement}, vol.~51, pp.~469--479, Jun
  2002.

\bibitem{Kanna_Stabilising_Clarke_El_Letters_2017}
{S. Kanna and D. P. Mandic}, ``Self-stabilising adaptive three-phase transforms
  via widely linear modelling,'' {\em Electronics Letters}, vol.~53, no.~13,
  pp.~875--877, 2017.

\bibitem{Huang_SE_SPM2012}
Y.~F. Huang, S.~Werner, J.~Huang, N.~Kashyap, and V.~Gupta, ``State estimation
  in electric power grids: {M}eeting new challenges presented by the
  requirements of the future grid,'' {\em IEEE Signal Processing Magazine},
  vol.~29, pp.~33--43, Sept 2012.

\bibitem{Microsemi}
``{Park, Inverse Park and Clarke, Inverse Clarke Transformation: MSS Software
  Implementation},'' Tech. Rep. 50200359-0/11.13, Microsemi Corporation, Aliso
  Viejo, California, January 2013.

\bibitem{Kay93}
S.~M. Kay, {\em Fundamentals of Statistical Signal Processing: Estimation
  Theory}.
\newblock Prentice Hall International, Inc, 1993.

\bibitem{Park_1929}
R.~H. Park, ``Two-reaction theory of synchronous machines generalized method of
  analysis -- {Part I},'' {\em Transactions of the American Institute of
  Electrical Engineers}, vol.~48, pp.~716--727, July 1929.

\bibitem{Shlens_PCA2014}
J.~Shlens, ``A tutorial on principal component analysis,'' {\em CoRR},
  vol.~abs/1404.1100, 2014.

\bibitem{fortescue1918method}
C.~L. Fortescue, ``Method of symmetrical co-ordinates applied to the solution
  of polyphase networks,'' {\em Transactions of the American Institute of
  Electrical Engineers}, vol.~37, no.~2, pp.~1027--1140, 1918.

\bibitem{Paap_Symmetrical_2000}
G.~C. Paap, ``Symmetrical components in the time domain and their application
  to power network calculations,'' {\em IEEE Transactions on Power Systems},
  vol.~15, pp.~522--528, May 2000.

\bibitem{Mandic_Book}
D.~P. Mandic and V.~S.~L. Goh, {\em Complex valued nonlinear adaptive filters:
  {Noncircularity}, widely linear and neural models}.
\newblock Wiley, 2009.

\bibitem{schreier2010book}
P.~J. Schreier and L.~L. Scharf, {\em Statistical signal processing of
  complex-valued data: {The} theory of improper and noncircular signals}.
\newblock Cambridge University Press, 2010.

\bibitem{Bolen_Dip_Characterisation_2000}
L.~Zhang and M.~H.~J. Bollen, ``Characterisation of voltage dips in power
  systems,'' {\em IEEE Transactions on Power Delivery}, vol.~15, no.~2,
  pp.~827--832, 2000.

\bibitem{Bollen_Sags_1997}
M.~Bollen, ``Characterisation of voltage sags experienced by three-phase
  adjustable-speed drives,'' {\em IEEE Transactions on Power Delivery},
  vol.~12, no.~4, pp.~1666--1671, 1997.

\bibitem{Bollen_SagsElsevier_2003}
M.~Bollen and L.~Zhang, ``Different methods for classification of three-phase
  unbalanced voltage dips due to faults,'' {\em Electric Power Systems
  Research}, vol.~66, no.~1, pp.~59 -- 69, 2003.

\bibitem{Sithan_TSIPN_2015}
S.~Kanna, D.~H. Dini, Y.~Xia, S.~Y. Hui, and D.~P. Mandic, ``Distributed widely
  linear {K}alman filtering for frequency estimation in power networks,'' {\em
  IEEE Transactions on Signal and Information Processing over Networks},
  vol.~1, pp.~45--57, March 2015.

\bibitem{Xia_3Point_2014}
Y.~Xia, Z.~Blazic, and D.~Mandic, ``Complex-valued least squares frequency
  estimation for unbalanced power systems,'' {\em IEEE Transactions on
  Instrumentation and Measurement}, vol.~PP, no.~99, pp.~1--1, 2014.

\bibitem{Xia_AMVDRGrid_2013}
Y.~Xia and D.~Mandic, ``Augmented {MVDR} spectrum-based frequency estimation
  for unbalanced power systems,'' {\em IEEE Transactions on Instrumentation and
  Measurement}, vol.~62, no.~7, pp.~1917--1926, 2013.

\bibitem{Picinbono_WL95}
B.~Picinbono and P.~Chevalier, ``Widely linear estimation with complex data,''
  {\em IEEE Trans. Signal Process.}, vol.~43, no.~8, pp.~2030--2033, 1995.

\bibitem{Akke_FM_1997}
M.~Akke, ``Frequency estimation by demodulation of two complex signals,'' {\em
  IEEE Transactions on Power Delivery}, vol.~12, pp.~157--163, Jan 1997.

\bibitem{Annette_unbalance2001}
A.~von Jouanne and B.~Banerjee, ``Assessment of voltage unbalance,'' {\em IEEE
  Trans. Power Del}, vol.~16, pp.~782--790, Oct 2001.

\bibitem{Clarke_Book}
E.~Clarke, {\em Circuit Analysis of A.C. Power Systems}.
\newblock New York: Wiley, 1943.

\bibitem{Picinbono94}
B.~Picinbono, ``{On Circularity},'' {\em IEEE Transactions on Signal
  Processing}, vol.~42, no.~12, pp.~3473--3482, 1994.

\bibitem{Xia_Balancing_2014}
Y.~Xia, K.~Wang, W.~Pei, and D.~P. Mandic, ``A balancing voltage transformation
  for robust frequency estimation in unbalanced power systems,'' in {\em Proc.
  of the Asia Pacific Signal and Information Processing Association Annual
  Summit and Conference (APSIPA)}, pp.~1--6, Dec 2014.

\bibitem{Javidi_ACLMS_08}
S.~Javidi, M.~Pedzisz, S.~L. Goh, and D.~P. Mandic, ``The augmented complex
  least mean square algorithm with application to adaptive prediction
  problems,'' {\em Proc. 1st IARP Workshop on Cognitive Information
  Processing}, pp.~54--57, 2008.

\end{thebibliography}

\end{document}